\documentclass[aip,jcp,superscriptaddress,amsmath,amssymb,twocolumn,reprint]{revtex4-1}

\usepackage[version=3]{mhchem} % Formula subscripts using \ce{}
\usepackage{graphicx}
\usepackage{amsmath}
\usepackage{amssymb}
\usepackage{nicefrac}
\usepackage{booktabs}

\usepackage{array}
\newcolumntype{m}{>{$} c <{$}}
\usepackage{color}

\def\rv{{\bf r}}
\def\fv{{\bf f}}

\def\beq{\begin{equation}}
\def\eeq{\end{equation}}

\begin{document} 

\author{Stefan Vuckovic}
\affiliation
{Department of Theoretical Chemistry and Amsterdam Center for Multiscale Modeling, FEW, Vrije Universiteit, De Boelelaan 1083, 1081HV Amsterdam, The Netherlands}
\author{Tom J. P. Irons}
\affiliation
{School of Chemistry, University of Nottingham, University Park, Nottingham NG7 2RD, United Kingdom}
\author{Lucas O. Wagner}
\affiliation
{Department of Theoretical Chemistry and Amsterdam Center for Multiscale Modeling, FEW, Vrije Universiteit, De Boelelaan 1083, 1081HV Amsterdam, The Netherlands} 
\author{Andrew M. Teale}
\affiliation
{School of Chemistry, University of Nottingham, University Park, Nottingham NG7 2RD, United Kingdom}
\author{Paola Gori-Giorgi}
\affiliation
{Department of Theoretical Chemistry and Amsterdam Center for Multiscale Modeling, FEW, Vrije Universiteit, De Boelelaan 1083, 1081HV Amsterdam, The Netherlands}
\email{p.gorigiorgi@vu.nl}

\title{Interpolated energy densities, correlation indicators and lower bounds from approximations to the strong coupling limit of DFT}
\begin{abstract}
We investigate the construction of approximated exchange-correlation functionals by interpolating locally along the adiabatic connection between the weak- and the strong-coupling regimes, focussing on the effect of using approximate functionals for the strong-coupling energy densities. The gauge problem is avoided by dealing with quantities that are all locally defined in the same way. Using exact ingredients at weak coupling we are able to isolate the error coming from the approximations at strong coupling only. We find that the nonlocal radius model, which retains some of the non-locality of the exact strong-coupling regime, yields very satisfactory results. We also use interpolation models and quantities from the weak- and strong-coupling regimes to define a correlation-type indicator and a lower bound to the exact exchange-correlation energy. Open problems, related to the nature of the local and global slope of the adiabatic connection at weak coupling, are also discussed.
\end{abstract}
%\begin{document}
\maketitle
%%%%%%%%%%%%%%%%%%%%%%%%%%%%%%%%%%%%%%%%%%%%%%%%%%%%%%%%%%%%%%%%%%%%%
%% The manuscript does not need to include \maketitle, which is
%% executed automatically.  The document should begin with an
%% abstract, if appropriate.  If one is given and should not be, the
%% contents will be gobbled.
%%%%%%%%%%%%%%%%%%%%%%%%%%%%%%%%%%%%%%%%%%%%%%%%%%%%%%%%%%%%%%%%%%%%%

\section{Introduction}

\label{sec_introedens}
Kohn--Sham Density Functional Theory (KS DFT)\cite{KohSha-PR-65} is the most widely used electronic structure method on account of its relatively low computational cost combined with an accuracy often rivalling that of much more expensive wavefunction--based methods. The success of KS DFT in any given application is however dependent on the quality of the approximation chosen to account for exchange--correlation (XC) effects. Since the original work of Kohn \& Sham,\cite{KohSha-PR-65} a wide range of functionals have been constructed to approximate the XC energy from local and semi--local quantities; in many cases, such XC functionals are sufficiently accurate that KS DFT may be used as a predictive tool in quantum chemistry. However there remain cases for which none of the presently available XC functionals provide an adequate approximation; an important example of this failure is in the treatment of strong correlation, commonly arising in systems exhibiting near--degeneracy.\cite{Sav-INC-96,CohMorYan-SCI-08,MorCohYan-PRL-09,CohMorYan-CR-12} The development of XC functionals that also perform well in challenging situations, such as bond dissociation, is essential to broaden the applications for which KS DFT works as a predictive tool.

In the previous work,\cite{locpaper} the construction of XC functionals able to treat strong and weak correlation effects with comparable accuracy has been explored through the approach of interpolating the local adiabatic connection (AC) between the weakly--interacting and strongly--interacting limits. It was observed\cite{locpaper} that the inclusion of both these limits offered a significant improvement in the treatment of strong (static) correlation without compromising the treatment of dynamical correlation. The accuracy of such interpolation schemes depends on two factors: the accuracy of the interpolation input parameters and the quality of the interpolation model itself. In the previous study,\cite{locpaper} interpolation models were tested with input parameters that had been computed to high accuracy, such that the quality of the model and the merits of interpolating the local (i.e., in each point of space) AC as opposed to the global (i.e., integrated over all space) AC could be assessed objectively. It was found\cite{locpaper} that the local AC interpolation generally yields more accurate XC energies than the global AC interpolation, for which these models were originally designed,\cite{locpaper,SeiPerLev-PRA-99,SeiPerKur-PRL-00} and are also more amenable to the construction of size consistent methods.

An essential consideration in the design of these local interpolation schemes\cite{locpaper} is the necessity to define all input quantities within a common \emph{gauge}. In contrast to global energies, local energy densities can differ by an arbitrary combination of spatial functions, since if their global integral is zero, the same total energy is obtained. In both the preceding and the present work, all quantities are defined within the gauge of the electrostatic potential of the XC hole,\cite{Bec-JCP-05,BecJoh-JCP-07,PerStaTaoScu-PRA-08,MirSeiGor-JCTC-12} namely the exchange energy density, slope of the local AC at the non--interacting limit and the XC energy density at the strongly--interacting limit.\cite{locpaper} Additionally, an approximation to the local initial slope of the AC has recently been derived; this approximation is a function of the occupied and virtual KS orbitals; it is in the same gauge as the exchange energy density, and is exact for two--electron systems.\cite{locpaper}

The structure of the exact strong coupling limit energy densities\cite{MirSeiGor-JCTC-12,VucWagMirGor-JCTC-15} provided by the \textit{strictly-correlated electrons} (SCE) theory\cite{Sei-PRA-99,SeiGorSav-PRA-07,GorVigSei-JCTC-09} is considerably different from that of the two input ingredients at weak correlation. Even if several authors have proposed different algorithms to compute them,\cite{MenLin-PRB-13,CheFriMen-JCTC-14,BenCarCutNenPey-arXiv-15,VucWagMirGor-JCTC-15} the SCE energy densities are still too expensive to compute and indeed are presently only available for relatively small systems.\cite{MirSeiGor-JCTC-12,VucWagMirGor-JCTC-15} It would seem therefore that, regardless of how well the local interpolation models perform, the computational hurdles of the SCE energy density precludes this approach from general use in DFT calculations. In the present work, means of overcoming this limitation are considered in the form of practical approximations to the energy densities in the strongly interacting limit, and the performance of the local interpolation models with the exact SCE quantities replaced with approximations is investigated. For this purpose, the well--established \textit{point-charge plus continuum} (PC) model\cite{SeiPerKur-PRA-00} and the more recent \textit{nonlocal radius} (NLR) model\cite{WagGor-PRA-14} are examined.

In addition to this, we use the interpolation approach to construct two quantities useful in the development of new XC functionals: a local indicator for the level of static correlation present, and a lower bound to the correlation energy that is tighter than those previously proposed.\cite{LieOxf-IJQC-81,MalGor-PRL-12}

\section{Theoretical Background}\label{sec_theo}     
\subsection{The adiabatic connection}\label{sec_ac}   
In this section we briefly review the adiabatic connection (AC) formalism\cite{HarJon-JPF-74,LanPer-SSC-75,GunLun-PRB-76,Har-PRA-84} and the quantities it defines in both global and local terms. The AC defines a link between the non--interacting KS auxiliary system and the physically--interacting system, comprising the ground state wavefunctions of the general Hamiltonian
\begin{equation}\label{eq:hamil_ac}
    \hat{H}_{\lambda} = \hat{T} + \lambda\hat{W} + \sum_{i} v_{\lambda}(\mathbf{r}_{i}),
\end{equation}
in which the electron interaction $\hat{W}$ is scaled by a coupling constant $\lambda$ and the one--body potential $v_{\lambda}$ varies such that the density is equal to that of the fully interacting system, $\rho_{\lambda=1}$, for all $\lambda \in \mathbb{R}$. Using the AC approach, we have an exact expression for the XC energy\cite{Har-PRA-84,LanPer-SSC-75}
\begin{align}
    E_{\rm xc}[\rho]= \int_{0}^{1} \mathcal{W}_{\lambda}[\rho]\,{\rm d}\lambda,
    \label{eq:ac_xc}
\end{align}
where $\mathcal{W}_{\lambda}[\rho]$ is the global AC integrand,
\begin{align}
    \mathcal{W}_{\lambda}[\rho] = \left\langle \Psi_{\lambda}[\rho]\left|\hat{W}\right|\Psi_{\lambda}[\rho]\right\rangle - U[\rho],
	\label{eq:w_lam}
\end{align} 
in which $\Psi_{\lambda}[\rho]$ is the ground state wavefunction of $\hat{H}_\lambda$ and $U[\rho]$ the Hartree (Coulomb) energy.

The global AC has been extensively studied and used to guide the construction of approximate XC functionals.\cite{SavColPol-IJQC-03,WuYan-JCP-03,PeaTeaToz-JCP-07,PeaMilTeaToz-JCP-08,TeaCorHel-JCP-09,TeaCorHel-JCP-10,TeaCorHel2-JCP-10} The first such functional was that proposed by Becke,\cite{Bec-JCP-93a} based on a model of the global AC, and many more have been developed since.\cite{Ern-CPL-96,CohMorYan-JCP-07,SeiPerLev-PRA-99,GorVigSei-JCTC-09,LiuBur-PRA-09} Most pertinent to the present work are those based on an interpolation of some form between the non--interacting and the strongly--interacting limits of the AC. The primary model of this class is the interaction strength interpolation (ISI) proposed by Seidl and co--workers\cite{SeiPerLev-PRA-99,SeiPerKur-PRL-00} and later revised in Ref.~\onlinecite{GorVigSei-JCTC-09}.

The expression for the XC energy given in eq.~\ref{eq:ac_xc} may be equivalently written as the spatial integral of a local quantity,
\begin{equation}\label{eq:ac_wxc}
    E_{\rm xc}[\rho] = \int_{0}^{1}{\rm d}\lambda\int \rho(\mathbf{r})\, w_{\lambda}(\mathbf{r}) {\rm d}\mathbf{r}
\end{equation}
in which $w_{\lambda}(\mathbf{r})$ is the energy density at coupling constant $\lambda$. As discussed in section~\ref{sec_introedens}, local quantities such as this are not uniquely defined hence any local interpolation scheme is only meaningful if constructed from quantities defined within the same gauge, which in this case is chosen to be that of the XC hole, defined as 
\begin{equation}\label{eq:wxc_def}
    w_{\lambda}(\mathbf{r}) = \frac{1}{2} \int \frac{h_{\rm xc}^{\lambda}(\mathbf{r},\mathbf{r}')}{|\mathbf{r} - \mathbf{r}'|} \,\mathrm{d} \mathbf{r}',
\end{equation}
where $h_{\rm xc}^{\lambda}(\mathbf{r},\mathbf{r}')$ is the XC hole,
\begin{equation}\label{eq_hxclambda}
    h_{\rm xc}^{\lambda}(\mathbf{r},\mathbf{r}') = \frac{P_{2}^{\lambda}(\mathbf{r},\mathbf{r}')}{\rho(\mathbf{r})}-\rho(\mathbf{r}').
\end{equation}
The pair--density $P_{2}^{\lambda}(\mathbf{r},\mathbf{r}')$ provides the link between the XC hole and the wavefunction $\Psi_{\lambda}[\rho]$ of eq.~\eqref{eq:w_lam} through the definition
\begin{equation}\label{eq:pair}\begin{aligned}
P_{2}^{\lambda}&(\mathbf{r},\mathbf{r}') = N(N-1) \times \\
& \sum_{\sigma_{1}\ldots\sigma_{N}} \int |\Psi_{\lambda}(\mathbf{r}\sigma_{1},\ldots,\mathbf{r}_{N}\sigma_{N})|^{2} \, \mathrm{d} \mathbf{r}_{3} \ldots \mathrm{d} \mathbf{r}_{N}.
\end{aligned}\end{equation}
The energy density may be locally integrated with respect to $\lambda$ to give the coupling--constant averaged (CCA) energy density,
\begin{equation}\label{eq:wxc_cca}
\bar{w}_{\rm xc}(\mathbf{r}) = \int_{0}^{1} w_\lambda(\mathbf{r}) \, \mathrm{d} \lambda,
\end{equation} 
in terms of which the XC energy may be written as
\begin{equation}\label{eq:exc_cca}
E_{\rm xc}[\rho] = \int \bar{w}_{\rm xc}(\mathbf{r})\rho(\mathbf{r})\,\mathrm{d}\mathbf{r}.
\end{equation} 

\subsection{AC interpolation models}\label{sec_acmodels}
In the previous study, the CCA energy density $\bar{w}_{\rm xc}(\mathbf{r})$ was approximated by interpolation models with near--exact quantities from the non--interacting and strongly--interacting limits.\cite{locpaper} As set out in section~\ref{sec_introedens}, these must be substituted by computationally inexpensive approximations for such an approach to yield practical density functional approximations (DFAs); this work assesses the suitability of several approximations to quantities in the strongly--interacting limit.

The interpolation models considered in this study are those which showed promise in prior work,\cite{locpaper} namely the model of Seidl, Perdew and Levy (SPL),\cite{SeiPerLev-PRA-99} the simplified model of Liu and Burke (LB)\cite{LiuBur-PRA-09} and the Pad\'e$[1/1]$ form,\cite{Ern-CPL-96,SanCohYan-JCP-06} all of which are summarized in appendix~\ref{app_interpol} for reference. Each requires three input quantities: $\mathcal{W}_{0}[\rho]$, $\mathcal{W}'_{0}[\rho]$ and $\mathcal{W}_{\infty}[\rho]$ for global interpolation, $w_{0}(\mathbf{r})$, $w'_{0}(\mathbf{r})$ and $w_{\infty}(\mathbf{r})$ for local interpolation. It is important to note that whilst both $\mathcal{W}_{\lambda}^{\rm LB}$ and $\mathcal{W}_{\lambda}^{\rm SPL}$ exhibit the correct behaviour in both the $\lambda \to 0$\cite{GorLev-PRB-93} and $\lambda \to \infty$\cite{GorVigSei-JCTC-09} limits, the corresponding Pad\'e$[1/1]$ model does not do so in the $\lambda \to \infty$ limit. 

In addition to these models, the following two--legged representation\cite{BurErnPer-CPL-97,locpaper} is also considered:
\begin{subequations}\begin{align}
w_{\lambda}(\mathbf{r}) &= 
\begin{cases}
w_{0}(\mathbf{r}) + \lambda w'_{0}(\mathbf{r}), & \lambda \leqslant x_{\rm corr} \\
w_{1}(\mathbf{r}), & \lambda > x_{\rm corr}
\end{cases} \label{eq:2leg} \\[2ex]
x_{\rm corr} &= \frac{w_{1}(\mathbf{r}) - w_{0}(\mathbf{r})}{w'_{0}(\mathbf{r})}. \label{eq:xlam}
\end{align}\end{subequations}
At the non--interacting limit, these models reduce simply to the exact exchange energy density $w_{0}(\mathbf{r})$, a key ingredient in local hybrids. The local initial slope $w'_{0}(\mathbf{r})$ has been studied in refs~\citenum{IroTea-MP-15} and~\citenum{locpaper}, whilst the energy density at the strongly--interacting limit $w_{\infty}(\mathbf{r})$\cite{MirSeiGor-JCTC-12,VucWagMirGor-JCTC-15,locpaper} is discussed in the following subsection.

\subsection{SCE energy densities}\label{sec_appx}
The SCE theory\cite{Sei-PRA-99,SeiGorSav-PRA-07,GorVigSei-JCTC-09} provides the framework in which exact global and local quantities can be obtained in the $\lambda \to \infty$ limit. The SCE energy density, in the gauge of the electrostatic potential of the XC hole, is given by\cite{MirSeiGor-JCTC-12,VucWagMirGor-JCTC-15,locpaper}
\begin{align}\label{eq:hxcsce}
    w_\infty(\rv) = \frac{1}{2} \sum_{k=2}^{N} \frac{1}{|\rv-\fv_k(\rv)|} - \frac{1}{2} v_{\rm H}(\rv),	
\end{align}
where $v_{\rm H}(\rv)$ is the Hartree potential and $\fv_k(\rv)$ are the \textit{co--motion functions}\cite{SeiGorSav-PRA-07,GorVigSei-JCTC-09,MirSeiGor-JCTC-12}. The co--motion functions parametrize the square of the $\Psi_\infty[\rho]$ wave function, which approaches a classical distribution in this limit\cite{SeiGorSav-PRA-07,GorVigSei-JCTC-09,MirSeiGor-JCTC-12}. For a given electron in the SCE system at position $\rv$, the positions of the other electrons will be determined by $\fv_k(\rv)$, which requires that several relations are satisfied as detailed in refs~\citenum{SeiGorSav-PRA-07} and~\citenum{MirSeiGor-JCTC-12}.

Due to the computational difficulties associated with the co--motion functions, SCE energy densities are only relatively easily computed for 1D\cite{MalMirGieWagGor-PCCP-14} and spherically symmetric systems.\cite{SeiGorSav-PRA-07,MirSeiGor-JCTC-12} To compute the SCE energy density for general 3D systems, one can invoke the dual Kantorovich SCE formulation,\cite{VucWagMirGor-JCTC-15} however computational cost still imposes a practical limitation on its use.

In considering approximations to the SCE energy density for use in local interpolation, there are several important properties to examine. The most important of these is that any such approximation must be defined in the gauge of the XC hole. Additionally models that give a reasonably accurate representation of the SCE energy densities, in particular having the correct asymptotic decay ($w_{\infty}(\mathbf{r}) \to -\frac{1}{2|\mathbf{r}|}$ as $|\mathbf{r}| \to \infty$), are more favourable. It is important to note however that more accurate $\mathcal{W}_{\infty}[\rho]$ and $w_{\infty}(\mathbf{r})$ does not necessarily result in a more accurately approximated $\bar{w}_{\rm xc}(\mathbf{r})$. This can occur where errors arising from inadequacies of the interpolation model partially cancel errors in the approximate input quantity, leading to a more accurate overall approximation. This will be further discussed in the following sections.

\subsection{NLR and PC energy densities}\label{sec_nlrpc}
In this subsection the approximations to the SCE functional used in this work, the NLR\cite{WagGor-PRA-14} and PC\cite{SeiPerKur-PRA-00} models,  are introduced and their respective properties discussed. 

The NLR functional has a non--local structure, inspired by the exact SCE functional,\cite{MirSeiGor-JCTC-12} with the model XC hole\cite{WagGor-PRA-14}
\begin{equation}\label{eq:hxcnlr}
    h_{\rm xc}^{\rm NLR}(\mathbf{r},\mathbf{r}') = -\rho(\mathbf{r}')\theta(r_{\rm NLR}(\mathbf{r})-\left|\mathbf{r}-\mathbf{r}'\right|),
\end{equation}
where $\theta(x)$ is the step function
\begin{equation}\label{eq:stepfun}
    \theta(x) = \begin{cases} 0 & x < 0 \\ 1 & x \geqslant 0 \end{cases} 
\end{equation}
and $r_{\rm NLR}(\mathbf{r})$ is the \textit{nonlocal radius}. Wagner and Gori-Giorgi defined $r_{\rm NLR}(\mathbf{r})$ by generalizing the Wigner--Seitz radius $r_{\rm s}$ to nonuniform densities\cite{WagGor-PRA-14}, satisfying the relation
\begin{equation}\label{eq:rnlr}
   \int \rho(\mathbf{r}')\theta(r_{\rm NLR}(\mathbf{r})-\left|\mathbf{r}-\mathbf{r}'\right|)\,{\rm d}\mathbf{r}' = 1,
\end{equation}
thus defining $r_{\rm NLR}(\mathbf{r})$ as the radius of a sphere containing one electron. This can equivalently be written as an integral over $\Omega(\mathbf{r})$, the volume of a sphere centred at $\mathbf{r}$ and with radius $r_{\rm NLR}(\mathbf{r})$,
\begin{equation}\label{eq:omeganlr}
   \int_{\Omega(\mathbf{r})} \rho(\mathbf{r}')\,{\rm d}\mathbf{r}' = 1.
\end{equation}
For systems with uniform densities, $r_{\rm NLR}(\mathbf{r})$ simply reduces to the Wigner--Seitz radius; a local function depending only on the density as $r_{\rm s}(\mathbf{r}) = [3/(4\pi\rho(\mathbf{r}))]^{1/3}$. For nonuniform systems, $r_{\rm NLR}(\mathbf{r})$ encodes nonlocal information, as it depends on the density at all the points in the sphere centred at $\mathbf{r}$ and with radius $r_{\rm NLR}(\mathbf{r})$.

From the model XC hole in eq.~\ref{eq:hxcnlr}, it can be shown\cite{WagGor-PRA-14} that the NLR energy density can be expressed as
\begin{equation}\label{eq:wnlr}
   w_{\infty}^{\rm NLR}(\mathbf{r}) = -\frac{1}{2}  \int_{\Omega(\mathbf{r})} \frac{\rho(\mathbf{r}')}{\left|\mathbf{r}-\mathbf{r}'\right|}\,{\rm d}\mathbf{r}'
\end{equation}
and thus the global equivalent $\mathcal{W}_{\infty}^{\rm NLR}[\rho]$ is given by\cite{WagGor-PRA-14}
\begin{equation}\label{eq:Wnlr}
   \mathcal{W}_{\infty}^{\rm NLR}(\mathbf{r}) = -\frac{1}{2} \int \int_{\Omega(\mathbf{r})} \frac{\rho(\mathbf{r})\rho(\mathbf{r}')}{\left|\mathbf{r}-\mathbf{r}'\right|}\,{\rm d}\mathbf{r}'\,{\rm d}\mathbf{r}.
\end{equation}
The NLR functional has been implemented in the \textsc{Gaussian}\cite{g09} and \textsc{Turbomole}\cite{TURBOMOLE} electronic structure packages.\cite{AntZhoErn-PRA-14,ZhoBahErn-JCP-15} Very recently, the NLR model has been also refined and improved by Bahman, Zhou \& Ernzerhof,\cite{BahZhoErn-JCP-16} with the addition of a shell of positive charge density, which is again inspired to the exact\cite{MirSeiGor-JCTC-12} SCE XC hole. The same authors have also shown how to implement both the original NLR model and their new shell model in a very efficient way.\cite{BahZhoErn-JCP-16} 

The other approximation for quantities in the strongly--interacting limit considered here is the PC model of Seidl and coworkers, in which $\lambda \to \infty$ quantities are modelled at a semilocal level.\cite{SeiPerKur-PRA-00} This approximation was developed before that many results on the exact SCE functional were available\cite{SeiGorSav-PRA-07,ButDepGor-PRA-12,MirSeiGor-JCTC-12} and, in contrast to the NLR approximation, does not model the XC hole directly. Detailed analysis does however show that the PC model, at least in its local density version of eq.~\eqref{eq:wpclda}, yields energy densities corresponding to the elecrostatic potential of the XC hole,\cite{MirSeiGor-JCTC-12} thus making them suitable for use in local interpolation schemes. Truncating the PC model at the LDA level leads to an energy density expressed as
\begin{equation}\label{eq:wpclda}
   w_{\infty}^{\rm PC-LDA}(\mathbf{r}) = -\frac{9}{10}  \left(\frac{4\pi}{3}\right)^{1/3}\rho(\mathbf{r})^{1/3},
\end{equation}
whilst truncating at the GGA level gives the energy density,
\begin{equation}\label{eq:wpcgga}\begin{aligned}
   w_{\infty}^{\rm PC-GGA}(\mathbf{r}) &= w_{\infty}^{\rm PC-LDA}(\mathbf{r}) \\
   &+ \frac{3}{350} \left(\frac{3}{4\pi}\right)^{1/3} \frac{|\nabla\rho(\mathbf{r})|^2}{\rho(\mathbf{r})^{7/3}}.
   \end{aligned}
\end{equation}

%%%%%%%%%%%%%%
\begin{figure}
\includegraphics[width=\linewidth]{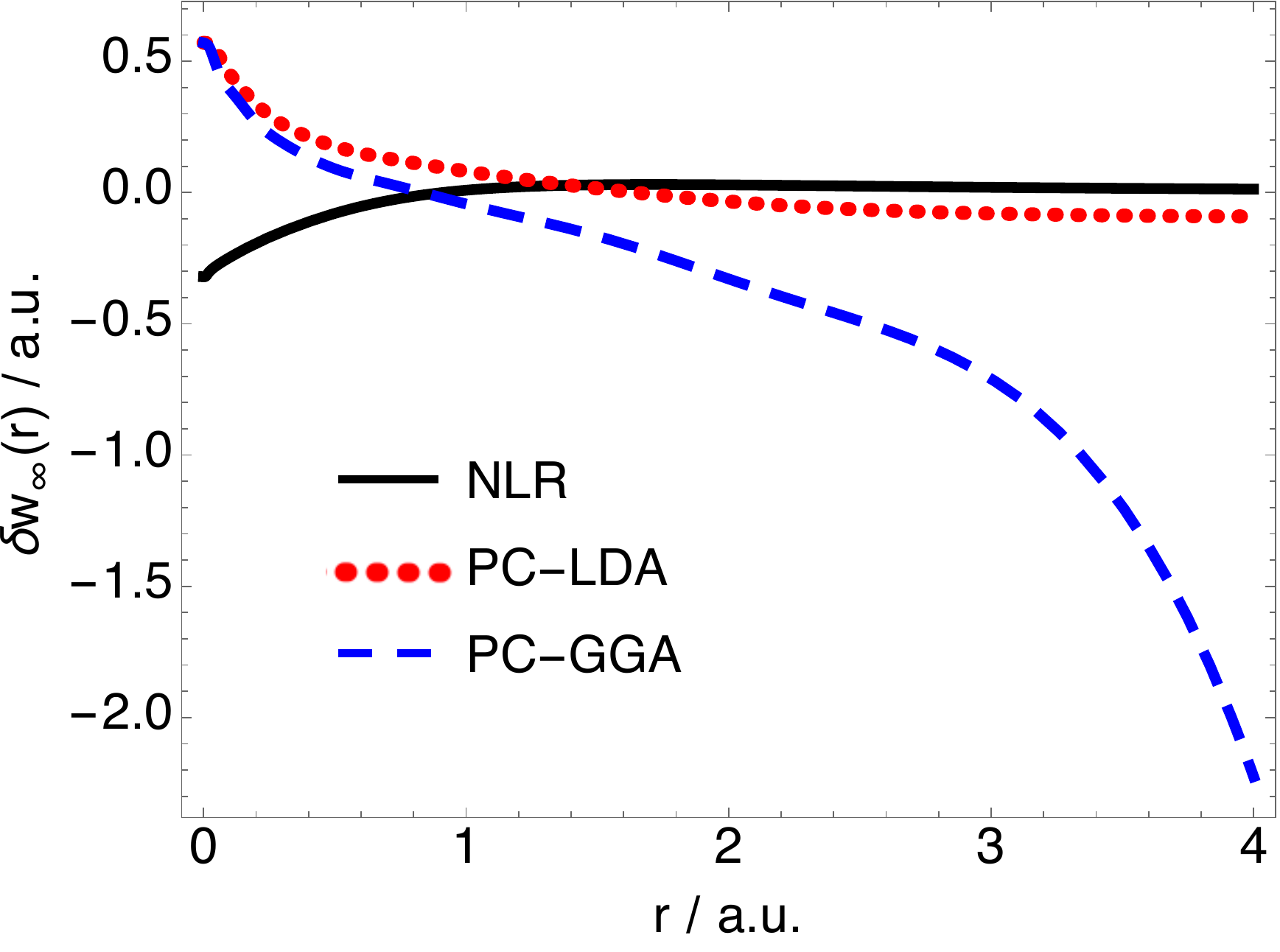}\\[2ex]
\includegraphics[width=\linewidth]{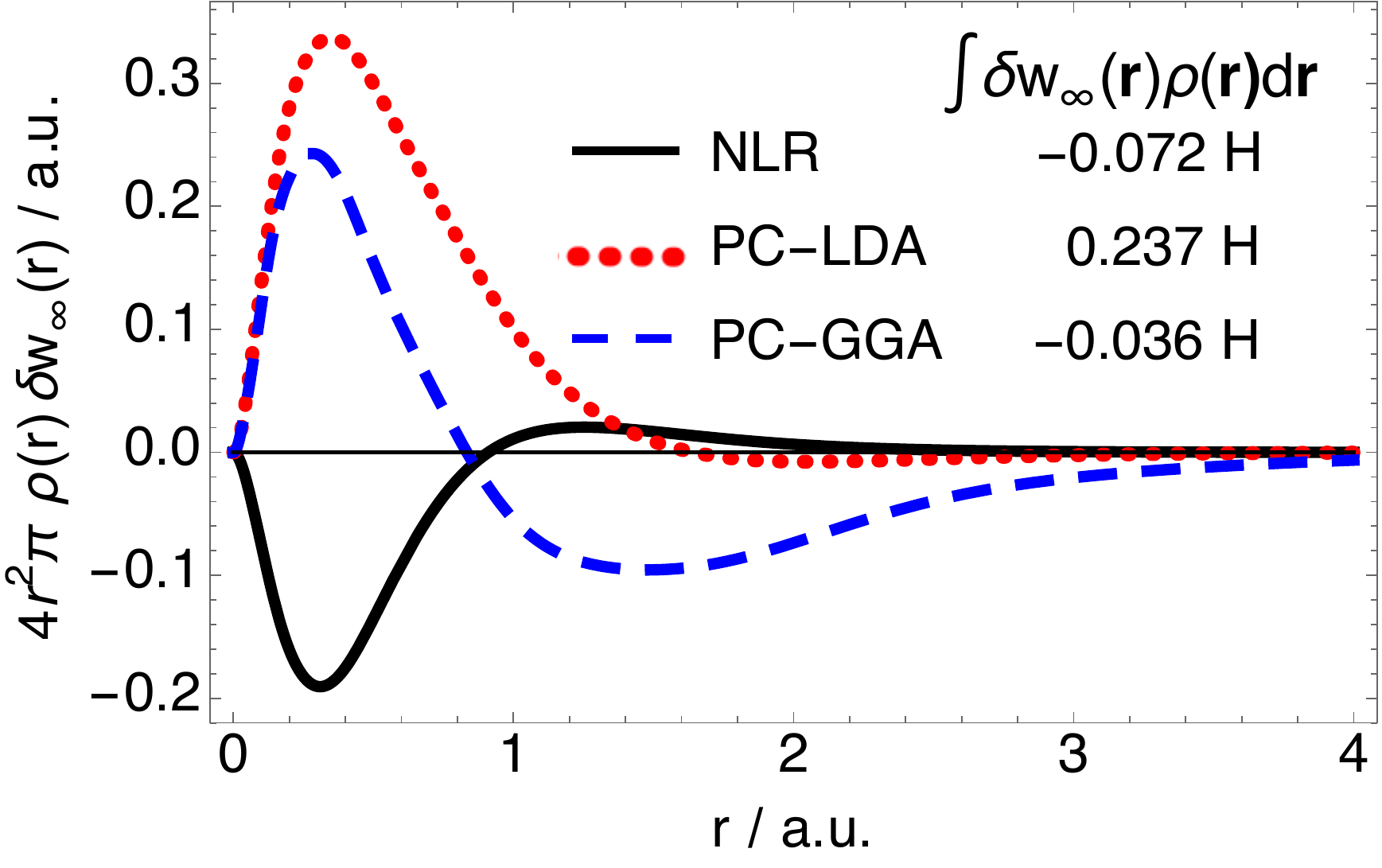}
\caption{Upper panel: Plots of the difference between the exact and approximate strong coupling limit energy density for the helium atom, $\delta w_\infty(\rv)=w_\infty(\rv)-w_\infty^{\rm model}(\rv)$, with respect to the distance from the nucleus, r / a.u. Lower panel: The quantity from the top panel multiplied by the density and spherical volume element.}
\label{fig_error_edens}
\end{figure}
%%%%%%%%%

In Figure~\ref{fig_error_edens}, the difference between $w_{\infty}(\mathbf{r})$ computed with the SCE theory and its approximate forms are plotted for the helium atom. It is evident that the NLR energy density is the most faithful to the SCE reference of the three, whilst the two PC forms present a slightly more nuanced picture. At the global level, the PC-GGA approximation appears to have an accuracy superior even to that of the NLR model, with the PC-LDA approximation yielding the greatest error by a considerable margin. However examination of the errors in the energy densities themselves reveals that the PC-GGA energy density exhibits a highly unphysical asymptotic decay, compared to the broadly reasonable asymptotic decay of the NLR and PC-LDA energy densities, suggesting that its apparent accuracy at the global level is largely the product of error cancellation\cite{MirSeiGor-JCTC-12}. 

Given the definition of the XC hole in eq.~\ref{eq_hxclambda}, and by virtue of the XC hole sum rule
\begin{equation}\label{eq_sumrule}
    \int h_{\rm xc}^{\lambda}(\mathbf{r},\mathbf{r}')\,{\rm d}\mathbf{r}' = -1,	
\end{equation}
it can be seen that $w_{\lambda}(\mathbf{r}) \to -\frac{1}{2|\mathbf{r}|}$ as $|\mathbf{r}| \to \infty$, for any given $\lambda$\cite{MirSeiGor-JCTC-12}. By virtue of eqs.~\ref{eq:wnlr} \&~\ref{eq:omeganlr}, the NLR energy density will exhibit the correct asymptotic behaviour. However, it is apparent from eqs.~\ref{eq:wpclda} \&~\ref{eq:wpcgga} that this will not be the case for the PC model energy densities.

%%%%%%%%%%
\begin{figure}
\includegraphics[width=\linewidth]{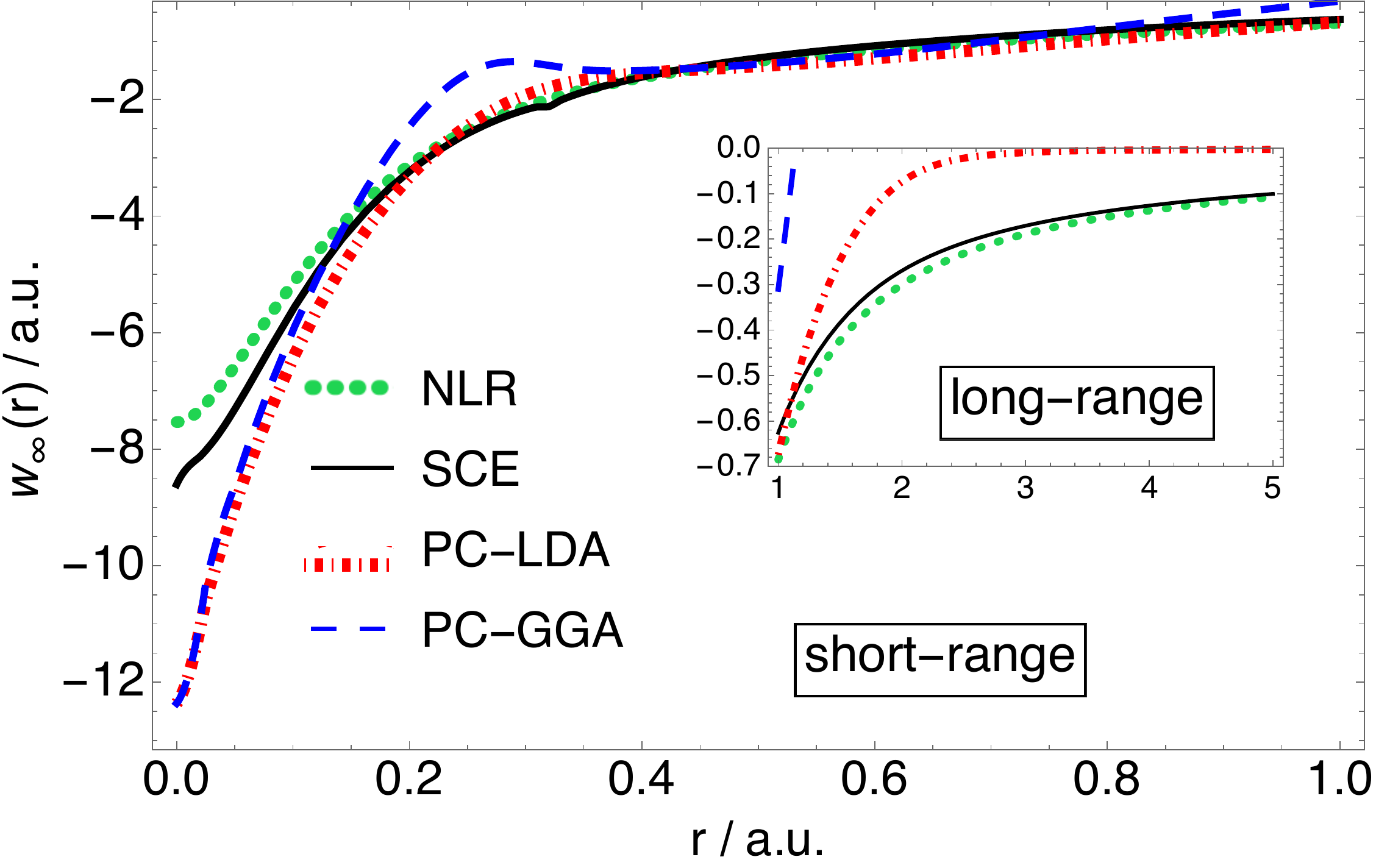}
\caption{The $w_\infty(\rv)$ energy densities in the Ne$^{6+}$ ion obtained by the following functionals: SCE, NLR, PC-LDA and PC-GGA.}
\label{fig_ne6dens}
\end{figure}
%%%%%%%%%

In Figure~\ref{fig_ne6dens} the SCE, NLR, PC-LDA and PC-GGA energy densities are plotted for the Ne$^{6+}$ ion (an example pertinent to later discussion). It can be seen that, as in the case of the beryllium atom discussed in ref.~\citenum{WagGor-PRA-14}, the NLR energy density lies above the SCE energy density in the core region and below in the valence region\cite{WagGor-PRA-14}. Additionally, it can be seen that the PC-GGA and PC-LDA energy densities have the predicted unphysical asymptotic behaviour, with the former even becoming positive at long range. 

This failing can present a challenge to the use of the PC models in local interpolation schemes as it causes both $w_{\infty}^{\rm PC-LDA}(\mathbf{r})$ and $w_{\infty}^{\rm PC-GGA}(\mathbf{r})$ to intersect $w_{0}(\mathbf{r})$. In the local SPL and LB schemes, the interpolated $w_{\lambda}(\mathbf{r})$ would have a non--zero imaginary component and thus be unphysical in regions where $w_{\infty}(\mathbf{r}) > w_{0}(\mathbf{r})$ as these models assume that $w_{\lambda}(\mathbf{r})$ monotonically decreases in $\lambda$. In contrast to the global AC, the monotonicity of the local AC (in the gauge of the XC hole) has not been formally proven; its assumption in these models is rationalised by the absence of known antithetic examples in Coulombic systems\cite{IroTea-MP-15,locpaper}. For practical purposes, the interpolation schemes were slightly adapted when using PC model approximations such that $w_{\lambda}(\mathbf{r}) \to w_{0}(\mathbf{r})$ where $w_{\infty}(\mathbf{r}) > w_{0}(\mathbf{r})$.

\section{Computational details}\label{sec_comp}
In this work, the Lieb maximization algorithm of refs.~\citenum{TeaCorHel-JCP-09,TeaCorHel-JCP-10} has been employed to compute accurate energy densities $w_{\lambda}(\mathbf{r})$ for $\lambda = 0$ and for a range of $\lambda \ll 1$, from which $w'_{0}(\mathbf{r})$ is numerically calculated by finite difference. Additionally, CCA energy densities $\bar{w}_{\rm xc}(\mathbf{r})$ are computed to provide accurate reference correlation energy densities\cite{IroTea-MP-15,locpaper}. The Lieb functional at interaction strength $\lambda$ is given by\cite{Lie-IJQC-83}
\begin{equation}\label{eq:lieb}
    F_{\lambda}[\rho] = \sup_{v} \left\lbrace E_{\lambda}[v] - \int v(\mathbf{r}) \rho(\mathbf{r}) \, {\rm d}\mathbf{r} \right\rbrace,
\end{equation}
in which $E_{\lambda}[v]$ is the energy yielded by a given electronic structure theory at potential $v(\mathbf{r})$, shown by Lieb to be the conjugate to the Lieb functional,
\begin{equation}\label{eq:liebEv}
    E_{\lambda}[v] = \inf_{\rho} \, \, \left\lbrace F_{\lambda}[\rho] + \int v(\mathbf{r}) \rho(\mathbf{r}) \, {\rm d}\mathbf{r} \right\rbrace.
\end{equation}

In the density--fixed AC formalism, the density is constrained to be equal to the $\lambda = 1$ density for all $\lambda$ by the optimizing potential $v(\mathbf{r})$. In this work, the potential is parameterized using the method of Wu and Yang (WY)\cite{WuYan-JCP-03} as
\begin{equation}\label{eq:wuyang}
    v(\mathbf{r}) = v_{\rm ext}(\mathbf{r}) + (1 - \lambda) v_{\rm ref}(\mathbf{r}) + \sum_{t} b_{t} g_{t}(\mathbf{r}),
\end{equation}
in which $v_{\rm ext}(\mathbf{r})$ is the external potential due to the nuclei, $v_{\rm ref}(\mathbf{r})$ a fixed reference potential which ensures the correct asymptotic behaviour of $v(\mathbf{r})$ and $\left\lbrace g_{t} \right\rbrace$ a set of Gaussian functions with $\left\lbrace b_{t} \right\rbrace$ their coefficients. Here, the Fermi--Amaldi potential\cite{Fermi1934} is used as the reference potential and the set of Gaussian functions $\left\lbrace g_{t} \right\rbrace$ is chosen to be the same as the orbital basis set, thus the Lieb functional is optimized with respect to the coefficients of the potential basis functions $\left\lbrace b_{t} \right\rbrace$. 

As in the preceding study,\cite{locpaper} Lieb maximisation calculations in this work were effected using the implementation of refs.~\citenum{TeaCorHel-JCP-09,TeaCorHel-JCP-10} in a development version of the \textsc{Dalton} quantum chemistry package\cite{Dalton-WIRES-14}, with the full configuration--interaction (FCI) and coupled--cluster singles and doubles (CCSD)\cite{Purvis1982} methods being used to compute $E_{\lambda}[v]$. All calculations of two--electron systems and those of LiH were performed with a FCI wave function, whilst the remaining systems were treated at the CCSD level. Additionally, the uncontracted aug-cc-pCVTZ basis set was selected as the orbital and potential basis set for all systems excluding LiH, for which the uncontracted cc-pVDZ basis was used instead\cite{Dun-JCP-89,WooDun-JCP-95}.

To compute SCE quantities for the atomic systems considered here, co--motion functions are obtained by using the conjectured SCE solution for spherically symmetric systems\cite{SeiGorSav-PRA-07}, yielding either exact or very accurate energy densities\cite{ColStr-arxiv-15,DiMGerNenSeiGor-xxx-15}. For the H$_2$ molecule, the SCE energy densities were computed via the dual Kantorovich SCE formulation\cite{ButDepGor-PRA-12,Dep-arxiv-15}, described in ref.~\citenum{VucWagMirGor-JCTC-15}. The key quantity necessary in the evaluation of the NLR energy densities is the NLR radius $r_{\rm NLR}(\mathbf{r})$, which was computed using the method described in ref.~\citenum{WagGor-PRA-14}. The PBE\cite{PerBurErn-PRL-96} and FCI dissociation curves for the hydrogen molecule were calculated using the \textsc{Dalton} quantum chemistry package \cite{Dalton-WIRES-14}.

\section{Results}\label{sec_results}
In this section, atomic correlation energies and H$_2$ dissociation curves computed using local interpolation models are presented, comparing the results obtained when using the SCE theory to model the strongly--interacting energy densities to those obtained using its NLR and PC approximations. As set--out earlier, the accurate data for the non--interacting energy density and the local initial slope\cite{locpaper} is used in the present evaluation of interpolation functionals as this allows the effect of substituting the SCE energy density with an approximation to be explicitly observed.  

We also report results on the LiH dissociation curve, in this case using always the NLR approximation for the energy densities in the strong correlation limit, comparing and rationalizing the performance of global and local interpolations.

\subsection{Atomic correlation energies}\label{sub_atoms}
The correlation energies of several atomic/ionic systems computed with the SPL and LB interpolation schemes and with the SCE, NLR and PC models providing the $\lambda \to \infty$ quantities are presented for global interpolation in Table~\ref{tab_glo} and local interpolation in Table~\ref{tab_loc}. Among these systems are two that are typically poorly described by contemporary DFAs; these are the H$^-$ ion, generally not predicted to be bound\cite{MirUmrMorGor-JCP-14} and the Ne$^{6+}$ ion which belongs to the beryllium isoelectronic series, a series exhibiting strong near--degeneracy effects with increasing nuclear charge $Z$.\cite{SavColPol-IJQC-03,locpaper}

It is evident from Tables~\ref{tab_glo}\&\ref{tab_loc} that the local interpolation gives always a lower mean absolute error (MAE) than the corresponding global interpolation. Indeed, the least accurate local interpolation (the LB model using $w_{\infty}^{\rm PC-LDA}(\mathbf{r})$) has a lower MAE than the most accurate global interpolation (the SPL model using $\mathcal{W}_{\infty}^{\rm NLR}[\rho]$).  

\begin{figure}
\includegraphics[width=\linewidth]{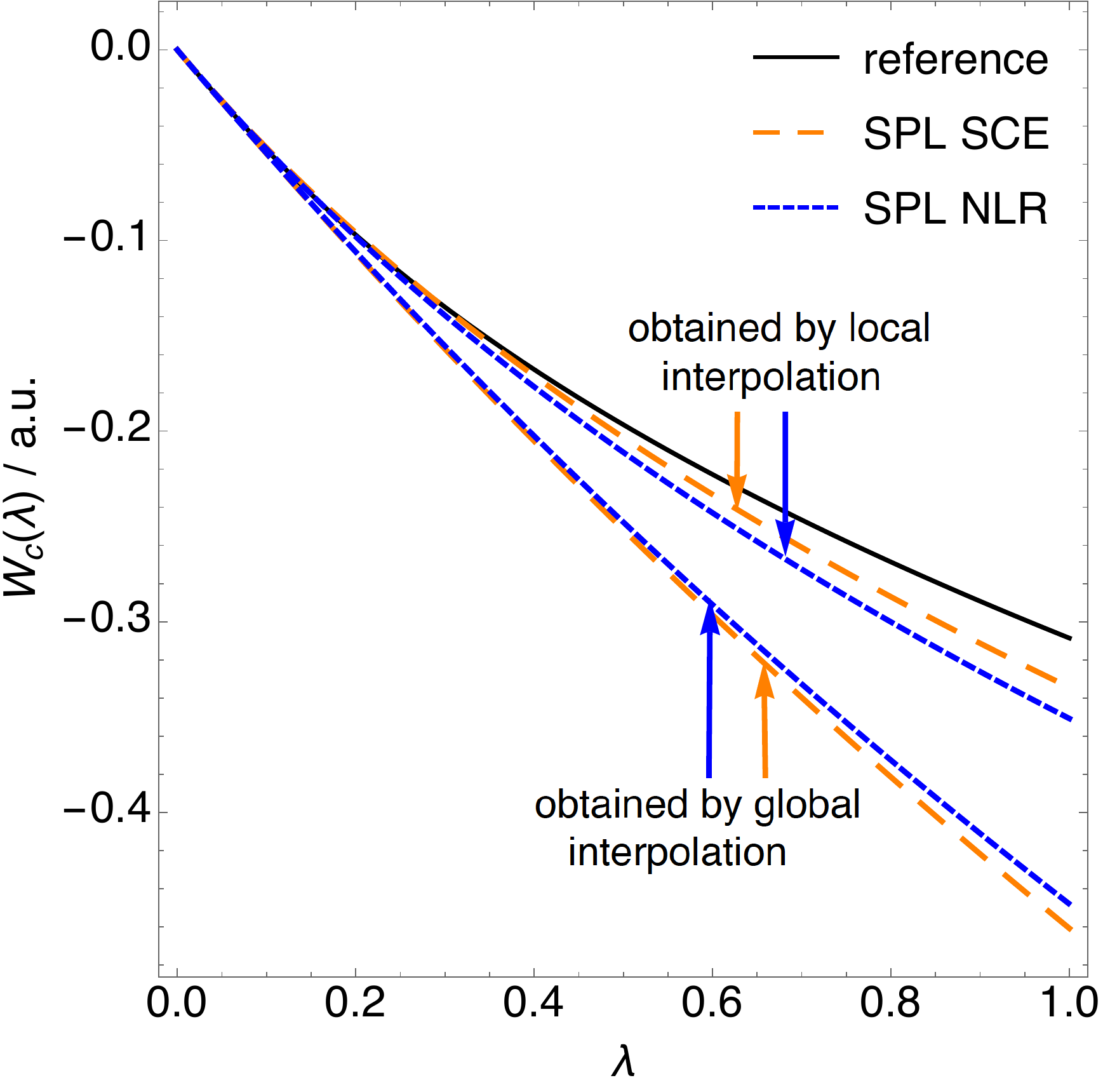}
\caption{The global correlation AC curves for the Ne$^{6+}$ ion obtained by the CCSD (reference) and by the global and local SPL interpolation with the SCE quantities and NLR approximation to the SCE quantities.}
\label{fig_gloacne6}
\end{figure}

In Figure~\ref{fig_gloacne6}, the global AC (with exchange omitted) obtained by SPL global and local interpolation with both the SCE and NLR input quantities are presented for the Ne$^{6+}$ ion. It can be seen that the curves pertaining to the local interpolation schemes are considerably closer to the reference AC (Lieb/CCSD) than those pertaining to global interpolation. This would indicate that advantage conferred by a local interpolation approach over a global interpolation approach is significantly greater than any depreciation in accuracy resulting from approximating SCE quantities with those of the NLR model. Furthermore, it can be seen from Table~\ref{tab_loc} that the accuracy of these atomic correlation energies depends more on the interpolation model chosen than the accuracy of the $\lambda \to \infty$ quantities. 

\begin{table*}
\caption{The atomic (ionic) correlation energies obtained by the global SPL and LB interpolation with the exact (SCE) and approximate (NLR, PC-LDA and PC-GGA) $\mathcal{W}_{\infty}$ interpolation quantities.} \label{tab_glo}
\centering
\begin{tabular*}{\textwidth}{l@{\extracolsep{\fill}}lllllllll} \hline\hline\noalign{\vskip 1ex}
Species   & $E_c^{\rm ref}$ & SPL SCE & SPL NLR & SPL PC-GGA & SPL PC-LDA & LB SCE  & LB NLR  & LB PC-GGA & LB PC-LDA \\[0.5ex]\hline\noalign{\vskip 0.5ex}
H$^-$     & -0.0409         & -0.0368 & -0.0352 & -0.0360    & -0.0415    & -0.0399 & -0.0383 & -0.0391   & -0.0444   \\
He        & -0.0400         & -0.0381 & -0.0371 & -0.0376    & -0.0401    & -0.0396 & -0.0388 & -0.0392   & -0.0413   \\
Be        & -0.0920         & -0.1049 & -0.1025 & -0.1042    & -0.1095    & -0.0925 & -0.1071 & -0.1085   & -0.1128   \\
Ne$^{6+}$ & -0.1833         & -0.2447 & -0.2399 & -0.2435    & -0.2527    & -0.2526 & -0.2487 & -0.2517   & -0.2590   \\
Ne        & -0.3470         & -0.3940 & -0.3840 & -0.3940    & -0.4010    & -0.4050 & -0.3970 & -0.4050   & -0.4100   \\
Ar        & -0.4040         & -0.4880 & -0.4810 & -0.4880    & -0.4910    & -0.4940 & -0.4890 & -0.4940   & -0.4960   \\
MAE (mH)  & ~~~~-               & ~~~35      & ~~~32      & ~~~35         & ~~~38         & ~~~37      & ~~~37      & ~~~39        & ~~~43        \\[1ex]\hline\hline
\end{tabular*}
\end{table*}

\begin{table*}
\caption{The atomic (ionic) correlation energies obtained by the local SPL and LB interpolation with the exact (SCE) and approximate (NLR, PC-LDA and PC-GGA) $w_\infty$ interpolation quantities.} \label{tab_loc}
\centering
\begin{tabular*}{\textwidth}{l@{\extracolsep{\fill}}lllllllll} \hline\hline\noalign{\vskip 1ex}
Species   & $E_c^{\rm ref}$ & SPL SCE & SPL NLR & SPL PC-GGA & SPL PC-LDA & LB SCE  & LB NLR  & LB PC-GGA & LB PC-LDA \\[0.5ex]\hline\noalign{\vskip 0.5ex}
H$^-$     & -0.0409         & -0.0367 & -0.0344 & -0.0364    & -0.0416    & -0.0398 & -0.0375 & -0.0393   & -0.0444   \\
He        & -0.0400         & -0.0378 & -0.0370 & -0.0347    & -0.0393    & -0.0394 & -0.0388 & -0.0359   & -0.0405   \\
Be        & -0.0920         & -0.0876 & -0.0904 & -0.0763    & -0.0925    & -0.1049 & -0.0955 & -0.0804   & -0.0973   \\
Ne$^{6+}$ & -0.1833         & -0.1919 & -0.1997 & -0.1653    & -0.2036    & -0.2045 & -0.2124 & -0.1760   & -0.2156   \\
Ne        & -0.3470         & -0.3830 & -0.3720 & -0.3770    & -0.3870    & -0.3960 & -0.3860 & -0.3900   & -0.3980   \\
Ar        & -0.4040         & -0.4450 & -0.4360 & -0.4350    & -0.4510    & -0.4590 & -0.4510 & -0.4940   & -0.4640   \\
MAE (mH)  & ~~~~-               & ~~~16      & ~~~14      & ~~~17         & ~~~18         & ~~~23      & ~~~21      & ~~~26        & ~~~25        \\[1ex]\hline\hline 
\end{tabular*}
\end{table*}
%%%%%%%%%%%%%%%%%%%%%%%%%%%%%%%%%%%%%%%%%%%%%%%%%%%%%%%%%%%%%%%%%%%%%%%%%%%%
%%%%%%%%%%%%%%%%%%%%%%%%%%%%%%%%%%%%%%%%%%%%%%%%%%

%%%%%%%%%%
%%%%%%%%%
In addition, it can also be seen that the MAE for the SPL interpolation is smaller than that for the LB interpolation scheme for both the global and local schemes. Interestingly, interpolating using the NLR approximation rather than the exact SCE form of $\mathcal{W}_\infty[\rho]$ and $w_\infty(\mathbf{r})$ results in a lower MAE. It has previously been observed that interpolation using SCE quantities often leads to an underestimation of atomic correlation energies\cite{locpaper} and that higher values of  $\mathcal{W}_\infty[\rho]$ results in a higher globally interpolated correlation energy. In these systems, $\mathcal{W}_\infty^{\rm NLR}[\rho] \geqslant \mathcal{W}_\infty[\rho]$ and thus has the effect of partially offsetting the interpolation error. Whilst correlation energies obtained by global interpolation with SCE are a lower bound to those obtained with the NLR model in its place, this is not necessarily the case for the local correlation schemes. As shown in Table~\ref{tab_loc}, for Be and Ne$^{6+}$ the correlation energies obtained by the interpolation with $w_\infty^{\rm NLR}(\mathbf{r})$ are lower than those obtained with the exact $w_\infty(\mathbf{r})$. This is reflected in the AC curves for Ne$^{6+}$, shown in Figure~\ref{fig_gloacne6}, in which that obtained by local interpolation with NLR lies below that resulting from local interpolation with SCE, despite the fact that $\mathcal{W}_\infty^{\rm NLR}[\rho] \sim -11.0~E_h$ compared to $\mathcal{W}_\infty[\rho]\sim -11.5~E_h$. 
%%%%%%%%%%
\begin{figure}
\includegraphics[width=\linewidth]{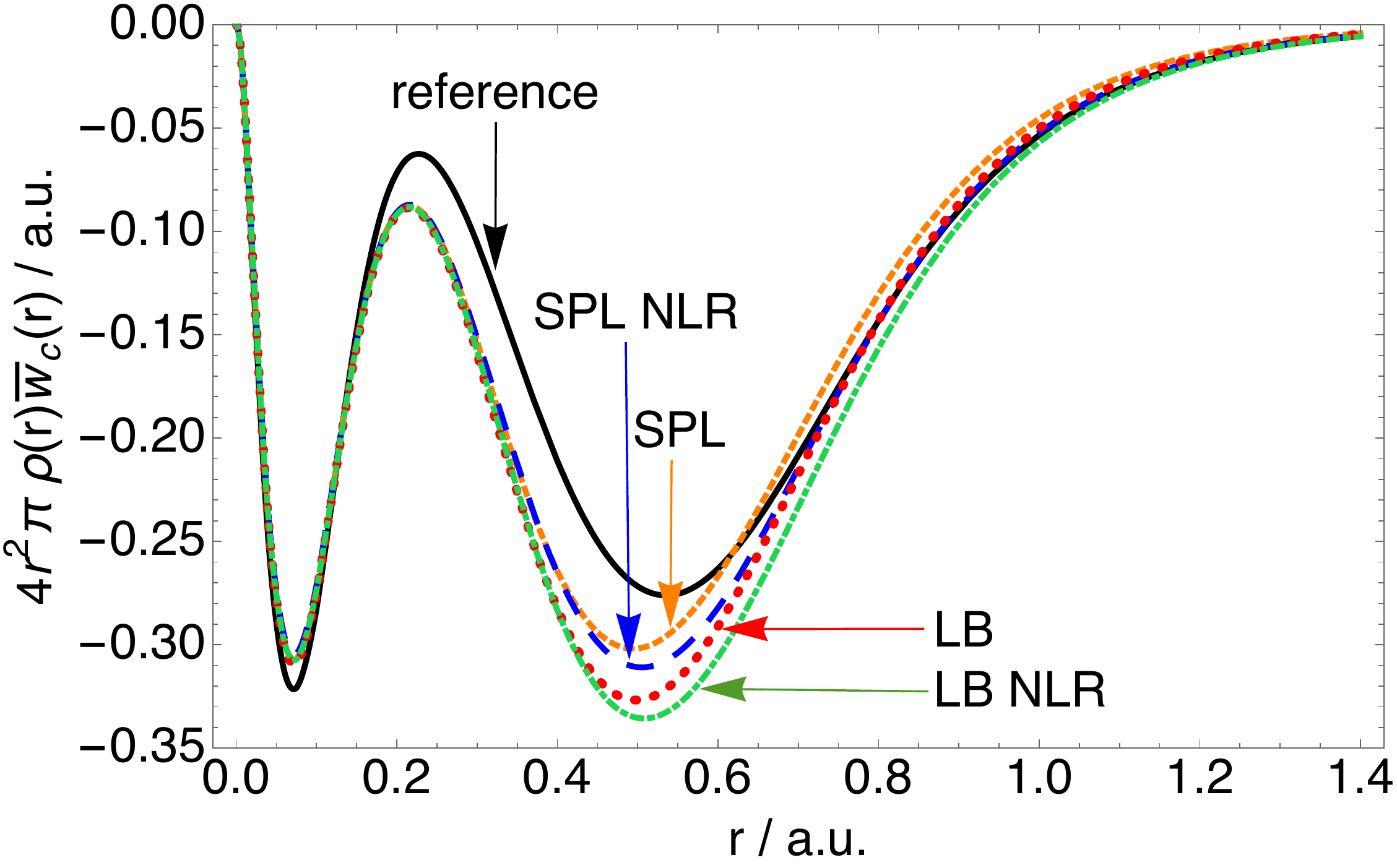}
\caption{The coupling constant averaged correlation energy densities multiplied by the density and spherical volume element for the Ne$^{6+}$ ion, obtained by both LB and SPL interpolation models and employing both the SCE (``SPL '' and ``LB'') and NLR $\lambda \to \infty$ energy densities (``SPL NLR'' and ``LB NLR''). The reference curve has been computed at the CCSD level.}
\label{fig_ne6}
\end{figure}
%%%%%%%%%
To rationalize these observations, it is useful to consider the corresponding energy densities; Figure~\ref{fig_ne6} shows the CCA correlation energy density in Ne$^{6+}$, multiplied by density and spherical volume element, obtained with both SPL and LB interpolation each using $w_\infty(\mathbf{r})$ and $w_\infty^{\rm NLR}(\mathbf{r})$ as input quantities, with the reference energy density for comparison. From this figure we can first notice that interpolation accuracy has a greater degree of dependence on interpolation model itself than on the accuracy of the $w_\infty(\mathbf{r})$ input parameter, as the energy densities exhibit a greater difference between SPL \& LB than between SCE \& NLR. We can also see that the two energy density curves that have been obtained by the interpolation with NLR quantities appear to be slightly below those obtained with the SCE quantities. As described in relation to Figure~\ref{fig_ne6dens}, in the core region of Ne$^{6+}$ ($r \lesssim 0.32$) $w_\infty^{\rm NLR}(\mathbf{r}) \geqslant w_\infty(\mathbf{r})$, whereas the opposite is generally true in the valence region. In the SPL and LB models, the sensitivity of the interpolated energy density to $w_\infty(\mathbf{r})$ is dependent on the magnitude of the local slope $w'_0(\mathbf{r})$; in regions where $w'_0(\mathbf{r}) \to 0$, $w_\lambda(\mathbf{r})$ simply approaches the exchange energy density as the correlation energy density itself vanishes, hence the accuracy of $w_\infty(\mathbf{r})$ has a minimal effect on the interpolated $w_\lambda(\mathbf{r})$. However, the converse is true in regions where $w'_0(\mathbf{r}) \to -\infty$, as the interpolated energy densities for both models would approach $w_\infty(\mathbf{r})$, making them highly sensitive to its accuracy. We can see in Figure~\ref{fig_ne6} how that reflects on the interpolated energy densities. In the core region, the NLR based interpolated energy densities (the ones labeled ``SPL NLR'' and ``LB NLR'') are hardly distinguishable from the ones that are based on the SCE (labeled ``SPL'' and ``LB''). Therefore, in the core region the interpolation neutralizes the difference between $w_\infty^{\rm NLR}(\rv)$ and $w_\infty(\rv)$. In the valence region, where the local interpolation is much more sensitive to the changes in $w_\infty(\rv)$, we can see that the NLR based interpolated energy densities are below the SCE based ones. This is why in the case of the Ne$^{6+}$ ion the NLR based local interpolation gives lower correlation energy than the SCE based local interpolation, exemplifying an interesting difference between the global and local interpolations.

\subsection{The H$_2$ dissociation curves}\label{sub_h2}
\begin{figure}
\includegraphics[width=\linewidth]{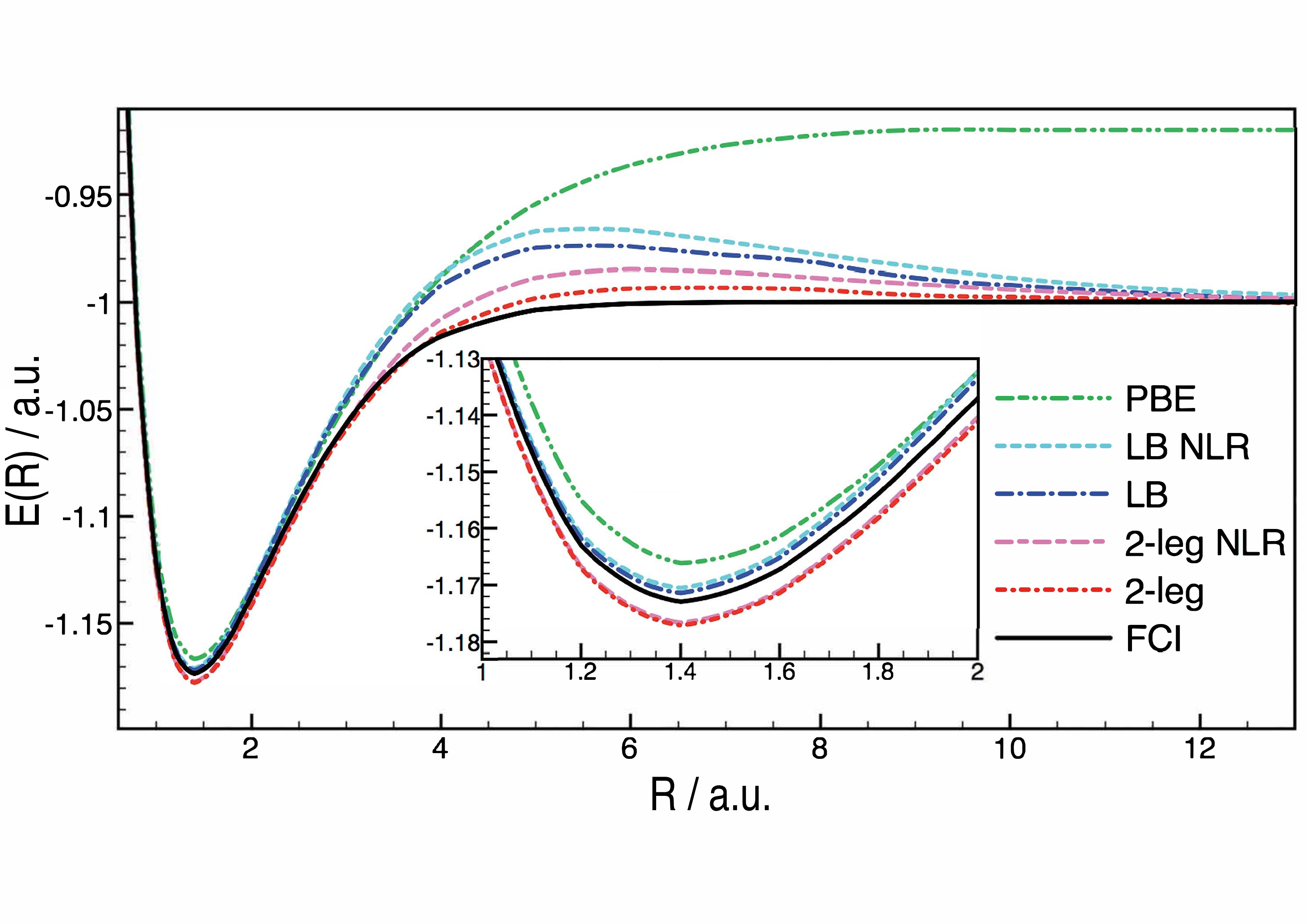}
\caption{The H$_2$ dissociation curves obtained by the local LB and two-legged representation interpolation with the SCE (``LB'' \& ``2-leg'') and NLR (``LB NLR'' \& ``2-leg NLR'') $w_\infty(\rv)$ interpolation input parameters. The PBE and FCI curves are shown for comparison.} 
\label{fig_h2curves}
\end{figure}

Figure~\ref{fig_h2curves} displays the H$_2$ dissociation curves obtained by local interpolation with the two--leg and LB models, using both SCE and NLR input parameters, in comparison to those acquired with FCI and the PBE functional (as described in section~\ref{sec_comp}).
%In Figure~\ref{fig_h2curves} we show the hydrogen molecule dissociation curves obtained by the local interpolation methods, comparing the NLR based local interpolation with the SCE based local interpolation. In the same figure we show the curves obtained by the FCI and the PBE functional (see the computational details in section~\ref{sec_comp}.) 

%%%%%%%%%%
%%%%%%%%%

It can be seen in Figure~\ref{fig_h2curves} that the dissociation curves given by the two interpolation models differ only slightly when the NLR model is used in place of the SCE model. Additionally, both models correctly predict the H$_2$ dissociation limit when using the NLR approximation, reflecting the underlying ability of the NLR model itself to dissociate the H$_2$ molecule correctly\cite{WagGor-PRA-14,locpaper}.

More pertinent however is the region of the dissociation curve at intermediate bond lengths, where DFT combined (or corrected with) methods that are constrained to be exact in the dissociation limit generally exhibits an unphysical ``bump'' (see, e.g., Refs.~\onlinecite{FucNiqGonBur-JCP-05,locpaper}). It was previously shown that the local interpolation approach, even with exact input parameters, does not entirely eliminate this erroneous feature, although it is significantly attenuated with the local two--legged interpolation model.\cite{locpaper} In Figure~\ref{fig_h2curves} we see that this feature is somewhat worsened when the NLR model is used in place of the SCE model for $w_\infty(\mathbf{r})$ as input for the local interpolation. In this region, the local initial slope is already large in magnitude (and increasing with bond length) thus there is a strong sensitivity of the interpolation models to the accuracy of $w_\infty(\mathbf{r})$. The approximate nature of the NLR model therefore has its most significant impact in this part of the dissociation curve. In contrast, approximating the SCE energy density with the NLR energy density in the interpolation models has considerably less effect on the energy computed at the equilibrium bond length, reflecting the lower sensitivity to the $\lambda \to \infty$ quantities. We also remark that all of the local interpolation forms are more accurate than spin-restricted PBE at the equilibrium bond length. The dissociation curves computed using the SPL model are omitted from Figure~\ref{fig_h2curves} for clarity, however it is noted that their properties are broadly similar to those of the LB curves but with a more pronounced unphysical feature at intermediate bond lengths\cite{locpaper}. 

\begin{figure}
\includegraphics[width=\linewidth]{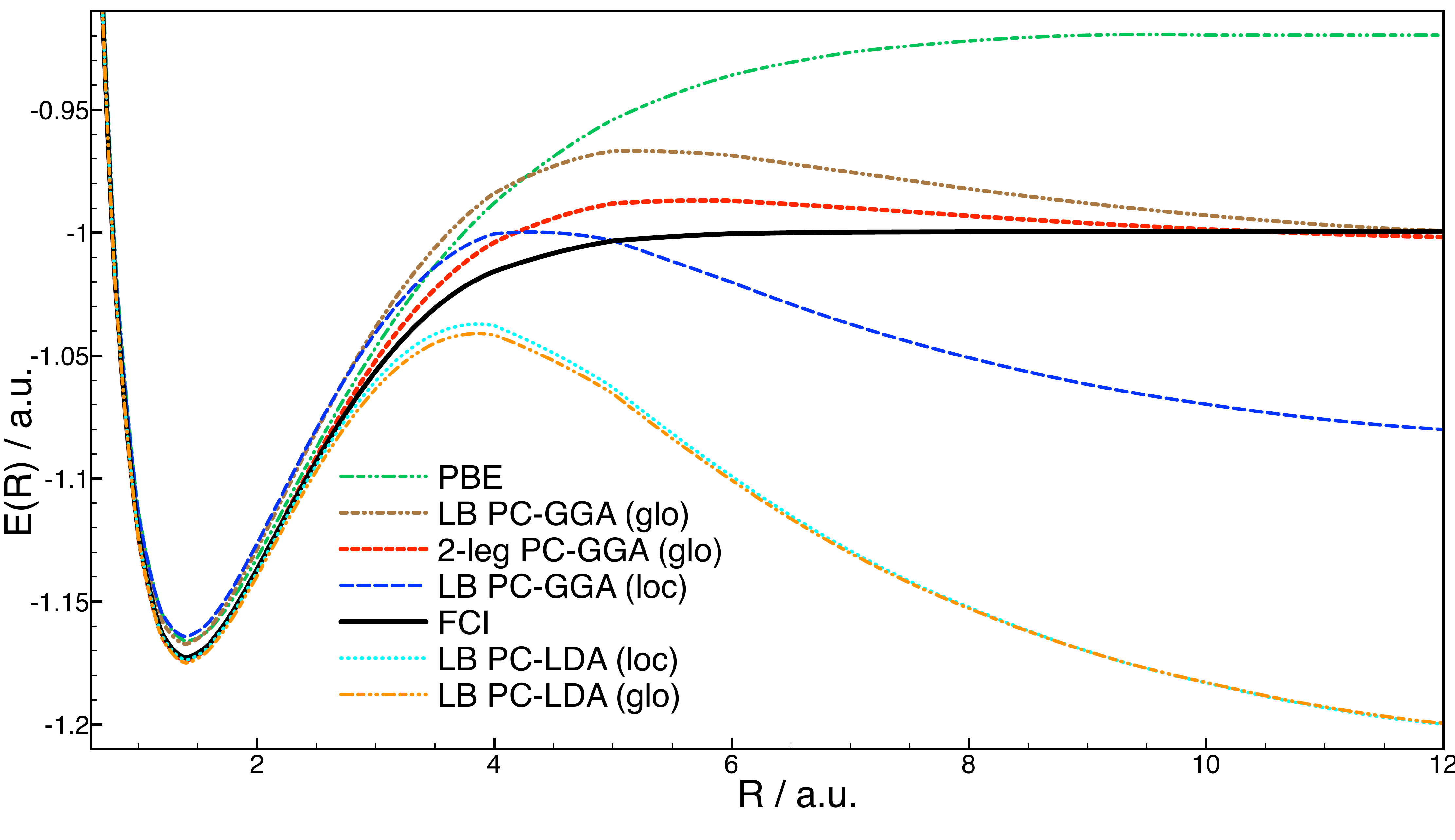}
\caption{H$_2$ dissociation curves obtained by local (``loc'') and global (``glo'') LB and two-legged interpolation models using PC-LDA and PC-GGA input parameters. The PBE and FCI curves are shown for comparison.}
\label{fig_h2curvespc}
\end{figure}
%%%%%%%%%

Figure~\ref{fig_h2curvespc} shows the H$_2$ dissociation curves obtained by interpolation using the PC model to approximate $w_\infty(\mathbf{r})$. As described in section~\ref{sec_appx}, in the present work the PC model is examined at both the LDA level and at the GGA level, containing gradient--dependent terms. It appears from Figure~\ref{fig_h2curvespc} that the most accurate dissociation curve obtained using PC model energy densities is the global two--legged model using PC-GGA to approximate $w_\infty(\mathbf{r})$. As observed in Figure~\ref{fig_h2curvespc}, the two--legged model gives a more accurate dissociation curve than the LB interpolation model, however in this case the differences between interpolation models themselves are overshadowed by those arising from the use of different $\lambda \to \infty$ quantities. There is also a marked difference between the dissociation curves obtained by global and local LB interpolation, using PC-GGA input quantities; whilst global interpolation yields a qualitatively accurate dissociation curve with only a small underestimation of the dissociation limit, local interpolation employing $w_\infty^{\rm PC-GGA}(\mathbf{r})$ yields a curve that becomes highly unphysical beyond the equilibrium geometry and results in an energy much lower than that of two hydrogen atoms in the dissociation limit. This is directly attributable to the superior performance of $\mathcal{W}_\infty^{\rm PC-GGA}[\rho]$ over $w_\infty^{\rm PC-GGA}(\mathbf{r})$ where, as seen in Figure~\ref{fig_ne6dens}, the PC-GGA energy density has erroneous long--range behaviour however global error cancellation results in a factitiously accurate $\mathcal{W}_\infty^{\rm PC-GGA}[\rho]$. In the local interpolation scheme, $w_\lambda(\mathbf{r}) \to w_0(\mathbf{r})$ where $w_\infty^{\rm PC}(\mathbf{r})$ crosses $w_0(\mathbf{r})$ and as such there is no equivalent error cancellation for local interpolation. 

The dissociation curves based on PC-LDA model input parameters are considerably poorer than those yielded by the PC-GGA model, both with a global and local scheme. For global interpolation, a quantitative comparison of their accuracies at the H$_2$ dissociation limit can be made by considering that the XC energy should cancel the Hartree energy and that $\mathcal{W}_0'[\rho] \to -\infty$ in this limit, hence $E_{\rm xc}[\rho] - U[\rho] = \mathcal{W}_\infty[\rho] - U[\rho] = 0$ should be satisfied. For the infinitely stretched H$_2$, $\mathcal{W}_\infty^{\rm PC}[\rho]$ will be twice $\mathcal{W}_\infty^{\rm PC}[\rho]$ evaluated on the density of a hydrogen atom. Whilst this error is relatively small for PC-GGA, $\mathcal{W}_\infty^{\rm PC-GGA}[\rho] - U[\rho] = -0.3~mE_h$, it is very large for PC-LDA, $\mathcal{W}_\infty^{\rm PC-LDA}[\rho] - U[\rho] = 312~mE_h$.

Whilst local interpolation using $w_\infty^{\rm NLR}(\mathbf{r})$ in place of the exact SCE quantities give reasonably accurate dissociation curves for H$_2$, those obtained with the computationally cheaper $w_\infty^{\rm PC}(\mathbf{r})$ appear volatile and unphysical and as such presently seem an inappropriate choice to substitute SCE energy densities in local interpolation models. The use of the PC model in global interpolation schemes appears to show more promise, yielding qualitatively accurate H$_2$ dissociation curves when using  $\mathcal{W}_\infty^{\rm PC-GGA}[\rho]$.\cite{MirSeiGor-JCTC-12} The accuracy of global interpolations that include the PC model and the issues coming from the lack of size consistency have been recently investigated in Ref.~\onlinecite{FabGorSeiSal-JCTC-16}.

\subsection{The lithium hydride dissociation curve}\label{sec_lih}
\begin{figure}
\includegraphics[width=\linewidth]{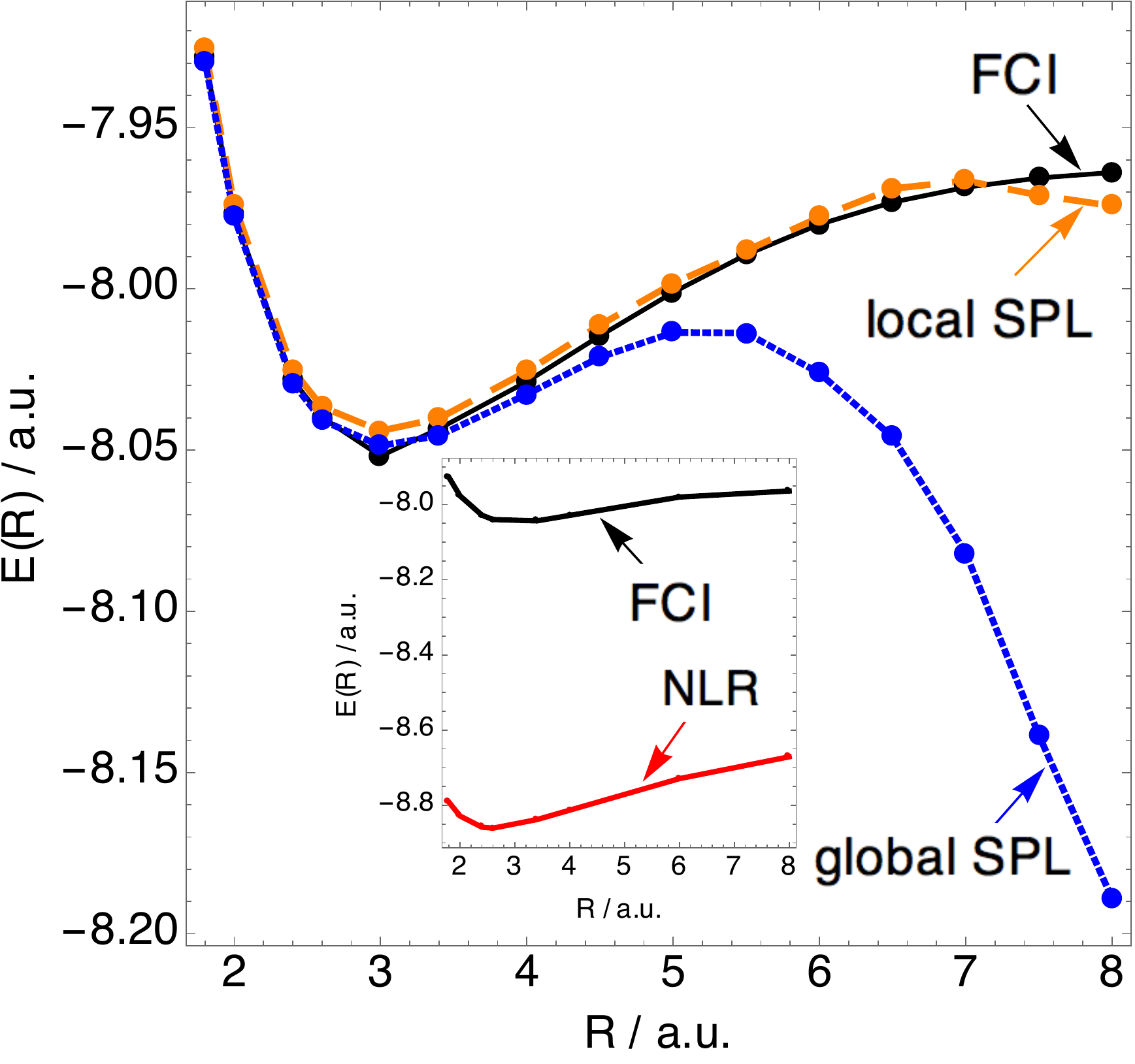}
\caption{The LiH dissociation curves obtained by FCI, NLR and the global and local SPL interpolation with the NLR quantities}
\label{fig_lih_cur}
\end{figure}

Size consistency within the global interpolation models that we use in this work is still preserved for systems that dissociate into equal fragments (assuming that the interpolation input quantities are size consistent themselves, which is a delicate issue for exchange and GL2 in a spin-restricted framework, see the discussion in Ref.~\onlinecite{FabGorSeiSal-JCTC-16}). For this reason, it would be interesting to compare the performance of the local interpolations against the global ones in the case of heterolytic dissociation. In Figure~\ref{fig_lih_cur} we show the  dissociation curves obtained by the local and global SPL interpolation and the reference (FCI) curve for comparasion. For both the global and local interpolation we used the approximate NLR input quantities: $\mathcal{W}_\infty^{\rm NLR}[\rho]$ and $w_\infty^{\rm NLR}(\rv)$, respectively.

%%%%%%%%%%
%%%%%%%%%

We can see that the energies of the stretched LiH obtained by the global SPL interpolation are unacceptably low. In the dissociation limit of LiH there is a step in the KS potential closer to the hydrogen, the more electronegative atom of the two.\cite{Per-DFM-85,TemMarMai-JCTC-09,GriBae-PRA-96} This step ensures that in the dissociation limit the atomic HOMO orbital energies are re-aligned and the molecule correctly dissociates into neutral atoms.\cite{Per-DFM-85,TemMarMai-JCTC-09,GriBae-PRA-96} This is also why the KS HOMO-LUMO gap closes in the dissociation limit of LiH, as it happens in the hydrogen molecule. As the gap closes and $\mathcal{W}'_0[\rho]$ diverges, the SPL globally interpolated $\mathcal{W}_\lambda[\rho]$ reduces to $\mathcal{W}_\infty^{\rm NLR}[\rho]$. The energy of the latter are extremely low as it can be seen from the inner panel of Figure~\ref{fig_lih_cur}, where we show the NLR dissociation curve that corresponds to the following approximation: $\mathcal{W}_\lambda[\rho] \approx \mathcal{W}_\infty^{\rm NLR}[\rho]$. 

In contrast to the globally interpolated, we can see that the locally interpolated energies are much more in-line with the reference data up to $R\sim 7$~a.u. However, even in this case, as the gap closes, the local slope will eventually tend to minus infinity everywhere,\cite{locpaper} making the locally-interpolated energies approach the NLR energies of the two fragments. The NLR energy is correct for the H atom (as NLR is exact for any one-electron density), but it overcorrelates the Li atom. We thus see that the local slope (even the exact one) is yet not suited to signal {\em locally} the amount of static correlation, as it is too dependent on a global quantity, namely the HOMO-LUMO gap. A possible way forward is to use other ideas to signal strong correlation such as those recently proposed in Refs.~\onlinecite{ZhoBahErn-JCP-15,KonPro-JCTC-15,YinRinPerSch-PRL-16} or to define a ``local gap'' as in Ref.~\onlinecite{GutSav-PRA-07}.

\section{Correlation indicator and lower bound to the total energy from an interpolation model}\label{sec_lower_ind}
In this section we use the two-legged representation interpolation model between weak and strong correlation to define and compute  a correlation type indicator (which, although it has very interesesting properties, does not address the issue discussed at end of the previous sec.~\ref{sec_lih}), and a lower bound to the exact energy, which is tighter than previously established ones.\cite{MalGor-PRL-12}.  

\subsection{Simple correlation type indicator}\label{cor_ind}
In DFT and quantum chemistry, electron correlation is usually classified into the intuitive concepts of dynamic and static (non-dynamic) correlation. Dynamic correlation is considered to be a short-range effect captured by perturbation theories such as MP2, which uses Hartree-Fock as a reference, or the G\"{o}rling-Levy (GL2) perturbation theory,\cite{GorLev-PRB-93} which uses the KS system as a reference. Static correlation, instead, is associated to near-degeneracy effects, with few or even a large number of important determinants in the physical description of the system at hand. These are cases in which a single determinantal reference is a bad starting point for perturbation theory. From the adiabatic connection point of view, a system dominated by dynamical correlation has a $\lambda$ dependence of $W_\lambda[\rho]$ that is very close to a straight line for $\lambda$ between 0 and 1, while a system with substantial static correlation has a $W_\lambda[\rho]$ that is substantially curved.
 
Burke, Ernzerhof and Perdew\cite{BurErnPer-CPL-97} have already noticed that the point of intersection between the two line segments used in their two-legged representation model has either a minimal (only static correlation) or a maximal (only dynamical correlation) value. We show here that the point of intersection of the two line segments in a different but similar two-legged representation,\cite{locpaper} is a very simple parameter that indicates the correlation type. This quantity is  $x_{\rm corr}(\rv)$ of eq~\ref{eq:xlam}. Notice that there are both local and global variants of this quantity. The local one is given in eq~\ref{eq:xlam}, whereas the global $X_{\rm corr}$ parameter is given by:
%	\begin{align}
%X_{\rm corr}=\frac{W_1-W_0}{W'_0}
%		\label{eq:gxlam}
%	\end{align}
\begin{equation}\label{eq:gxlam}
	X_{\rm corr} = \frac{\mathcal{W}_1-\mathcal{W}_0}{\mathcal{W}'_0}
\end{equation}
$X_{\rm corr}$ and $x_{\rm corr}(\rv)$ can have values between $0$ and $1$. If $X_{\rm corr}=1$, then the shape of the adiabatic connection integrand is linear for $\lambda$ values between $0$ and $1$ and GL2 captures all the correlation in the system. We can say that, if $X_{\rm corr}=1$, then the correlation in the system is purely dynamical. On the other hand, if $X_{\rm corr}=0$, then the adiabatic connection curve is L-shaped and all the correlation present in the system is static. 

%%%%%%%%%%
\begin{figure}
\includegraphics[width=\linewidth]{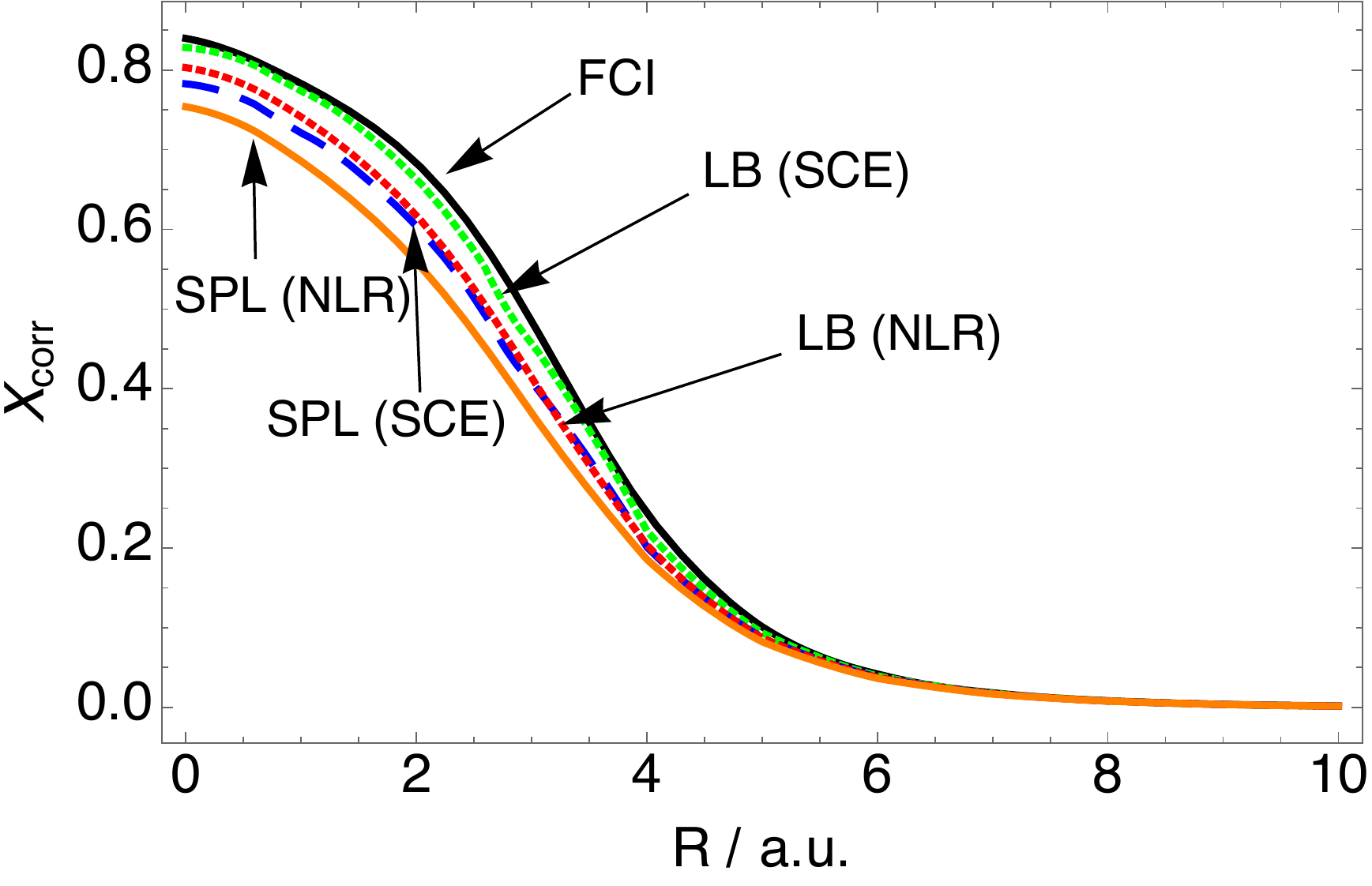}
\caption{$X_{\rm corr}$ correlation-type parameter of eq~\ref{eq:gxlam}, as a function of the H$_2$ bond length.}
\label{fig_xlamh2}
\end{figure}
%%%%%%%%%

In Figure~\ref{fig_xlamh2} we show the ``exact'' $X_{\rm corr}$ (the black curve) for the hydrogen molecule as a function of the bond length. At short bond lengths, the $X_{\rm corr}$ value is quite high, reflecting the dominance of dynamic correlation. Moreover, at short bond lengths it is expectedly very close to the $X_{\rm corr}$ value for the He atom ($X_{\rm corr} ~=~0.84$). As we stretch the H$_2$ bond, $X_{\rm corr}$ value decreases and finally drops to zero for the infinitely stretched H$_2$ in which all the present correlation is static. 

By virtue of eq~\ref{eq:gxlam}, we need the three input quantities to compute $X_{\rm corr}$: $\mathcal{W}_0[\rho]$, $\mathcal{W}'_0[\rho]$ and $\mathcal{W}_1[\rho]$. If we lack $\mathcal{W}_1[\rho]$ information, we can approximate this quantity by doing the interpolation that uses $\mathcal{W}_0[\rho]$, $\mathcal{W}'_0[\rho]$ and $\mathcal{W}_\infty[\rho]$ as input. For this purpose, we can employ the interpolation model that we use in this work, such as the SPL or LB model. In Figure~\ref{fig_xlamh2}, together with the ``exact'' $X_{\rm corr}$ for H$_2$ we show the approximate ones that have been calculated with interpolated $\mathcal{W}_1[\rho]$ from $\mathcal{W}_0[\rho]$, $\mathcal{W}'_0[\rho]$ and $\mathcal{W}_\infty[\rho]$. We used the two interpolation methods, SPL and LB and both the exact (SCE) and the NLR $\mathcal{W}_\infty[\rho]$ quantities. We can see in Figure~\ref{fig_xlamh2} that all the approximate curves follow the trend of the ``exact'' $X_{\rm corr}$ curve. In this case, the LB interpolation is more accurate than the SPL interpolation. What we also see is that $\mathcal{W}_1[\rho]$ interpolation is more sensitive to the interpolation form than the accuracy of the $\mathcal{W}_\infty$ quantity: the LB  $X_{\rm corr}$ curve based on $\mathcal{W}_\infty^{\rm NLR}[\rho]$ is even more accurate than the SPL $X_{\rm corr}$ curve that is based on the exact $\mathcal{W}_\infty[\rho]$.

Apart from indicating a type of correlation, the $X_{\rm corr}$  parameter can be very useful for telling us how accurate GL2 (or MP2) is for the given system. The closer the $X_{\rm corr}$ value is to $0$, the more poorly the two theories will describe correlation in the system. A better starting point for the correlation description in this case would be the KS SCE theory, which gives extremely low energies for the systems whose $X_{\rm corr}$ value is close to $1$. 

%%%%%%%%%%
\begin{figure}
\includegraphics[width=\linewidth]{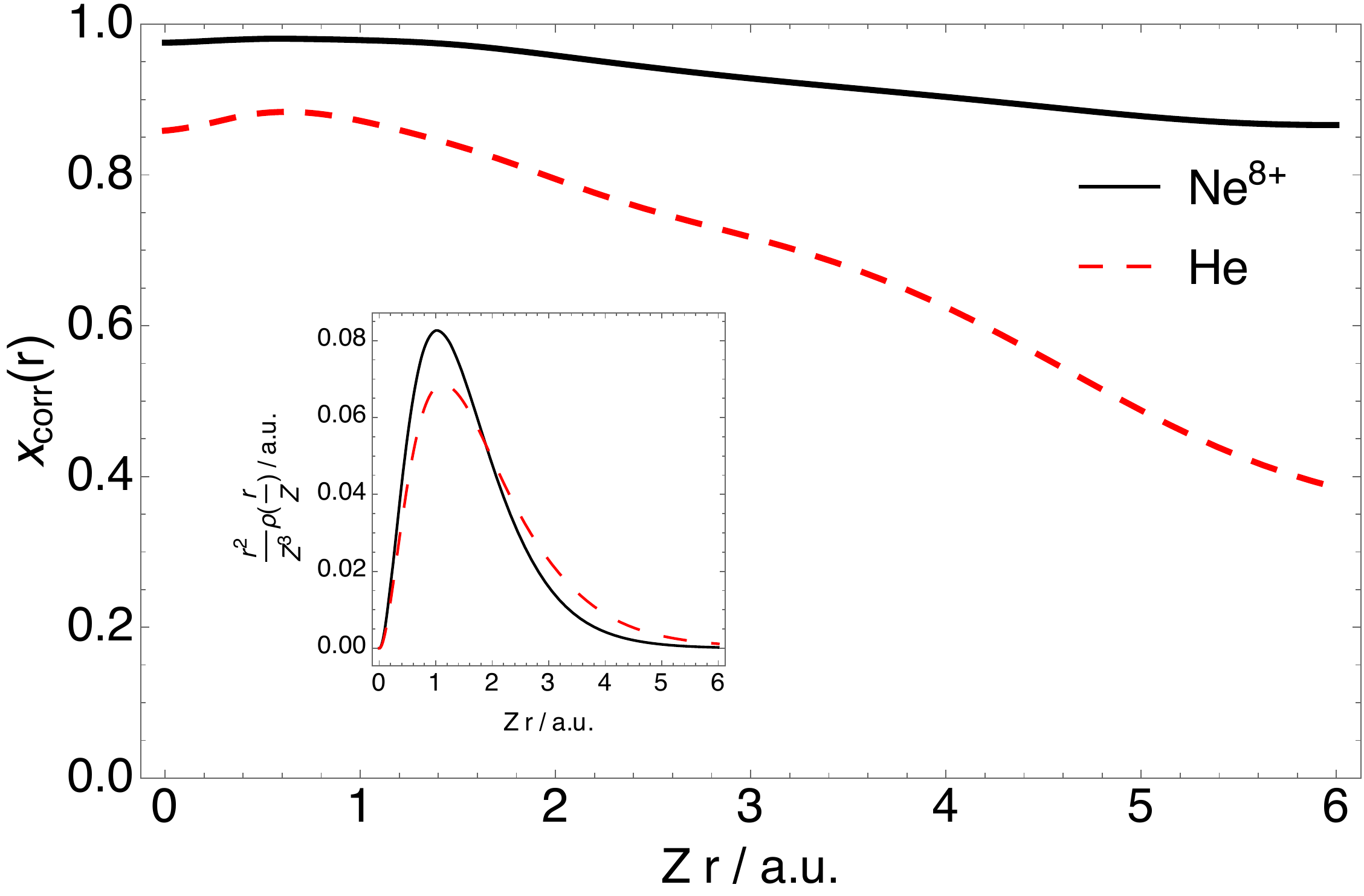}
\includegraphics[width=\linewidth]{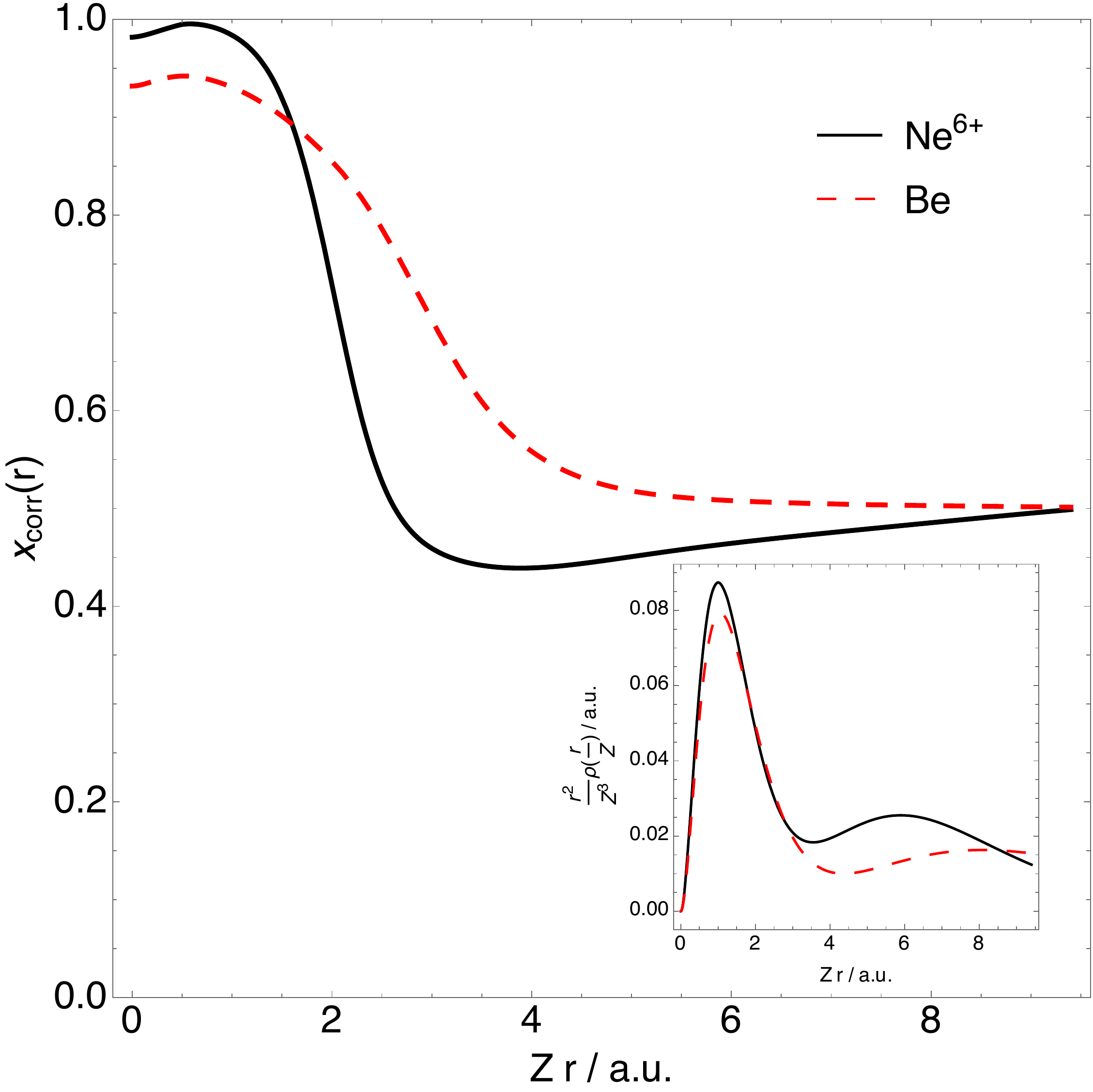}
\caption{$x_{\rm corr}(\rv)$ local correlation-type indicator of eq~\ref{eq:xlam}.  shown for the He atom and Ne$^{8+}$ ion atom (left panel) and for the Be atom and Ne$^{6+}$ ion (right panel), where r is a distance from the nucleus}
\label{fig_xlam_atoms}
\end{figure}
%%%%%%%%%

Grimme and Hansen\cite{GriHan-Ang-2015} have recently introduced a position--dependent indicator based on the fractional orbital occupation, aiming to detect molecular ``hot regions'' that have a high static correlation contribution. Notice that if we go from the $X_{\rm corr}$ value, which is a single number, to the $x_{\rm corr}(\rv)$ we get the correlation type indicator as a function of space. It is easier to visualize $x_{\rm corr}(\rv)$ than $w_\lambda(\rv)$, as the latter depends on both $\lambda$ and $\rv$. In the top panel of Figure~\ref{fig_xlam_atoms} we show $x_{\rm corr}(r)$ the He atom and the Ne$^{8+}$ ion, both belonging to the helium isoelectronic series. We can see that for the He atom the $x_{\rm corr}(r)$ decreases as we move from the nucleus, but in the energetically most important regions ($r \lesssim 2.0~a.u.$), $x_{\rm corr}(r)$ is quite high and also gives the high global $X_{\rm corr} = 0.84$ value. The $x_{\rm corr}(r)$ curve for the Ne$^{8+}$ ion is higher and flatter than that of He and the corresponding global $X_{\rm corr} = 0.97$ value is much closer to $1$. This indicates that dynamic correlation dominates even more in this ion than in the helium atom and this is what one would expect in the case of the helium isoelectronic series as the nuclear charge increases.\cite{locpaper} It is important to stress that in the asymptotic, low-density regions, the local indicator becomes too sensitive to numerical errors, since both numerator and denominator or eq~\ref{eq:xlam} become very small, so that it should not be trusted in these energetic unimportant regions.

In the bottom panel of Figure~\ref{fig_xlam_atoms} we show $x_{\rm corr}(r)$ for the Be atom and the Ne$^{6+}$ ion, members of the beryllium isoelectronic series. We can see that these two $x_{\rm corr}(r)$ curves exhibit almost a step structure, clearly distinguishing between the core region ($Z~r \lesssim 4.0~a.u.$ for Be and $Z~r \lesssim 3.2~a.u.$ for Ne$^{6+}$) and the valence region. In the core region the $x_{\rm corr}(r)$ value is high and close to $1.0$ for Ne$^{6+}$ ion and around $0.9$ for Be. This indicates the presence of almost purely dynamic correlation. The trend that the ion with a greater  $Z$ value has higher values of $x_{\rm corr}(r)$ still follows in this region. On the other hand, in the valence region $x_{\rm corr}(r)$  of the two curves is lower with the value about $0.5$, indicating a substantial contribution of static correlation and that is reflected by the rapid curvature of the  corresponding local AC curves.\cite{IroTea-MP-15}. Interestingly, the trend of two $x_{\rm corr}(r)$ curves is opposite in the valence region in which  $x_{\rm corr}(r)$ of Ne$^{6+}$ lies below that of Be. This observation signals that in the valence region of the former ion there is a higher contribution of static correlation. This can be understood in the light of the trend for KS HOMO-LUMO gap for the beryllium isoelectronic series, as the nuclear charge $Z$ increases.\cite{locpaper,SavColPol-IJQC-03} The KS HOMO-LUMO ($2s$-$2p$) gap of Be is actually smaller ($0.133~E_h$) than that of Ne$^{6+}$ ($0.481~E_h$). However, the absolute values of the Ne$^{6+}$ orbital energies are much higher. Therefore, for a fairer comparison we can use a relative KS orbital energy gap, which we define in the following way: $\frac{\epsilon_{2s}-\epsilon_{2p}}{\epsilon_{2s}}$. The value of the relative gap defined this way for Ne$^{6+}$ is much lower than that of Be: $0.070$ and $0.364$, respectively. The reported KS energy gaps have been calculated performing the maximisation of eq~\ref{eq:lieb} at $\lambda=0$ and using CCSD/aug-cc-pCVTZ level of theory for obtaining $E_0[v]$ (see section~\ref{sec_comp} for the details). They also correspond to the KS potential, which goes to zero as the distance from the nuclear charge goes to infinity.  
%%%%%%%%%%
\begin{figure}
\includegraphics[width=\linewidth]{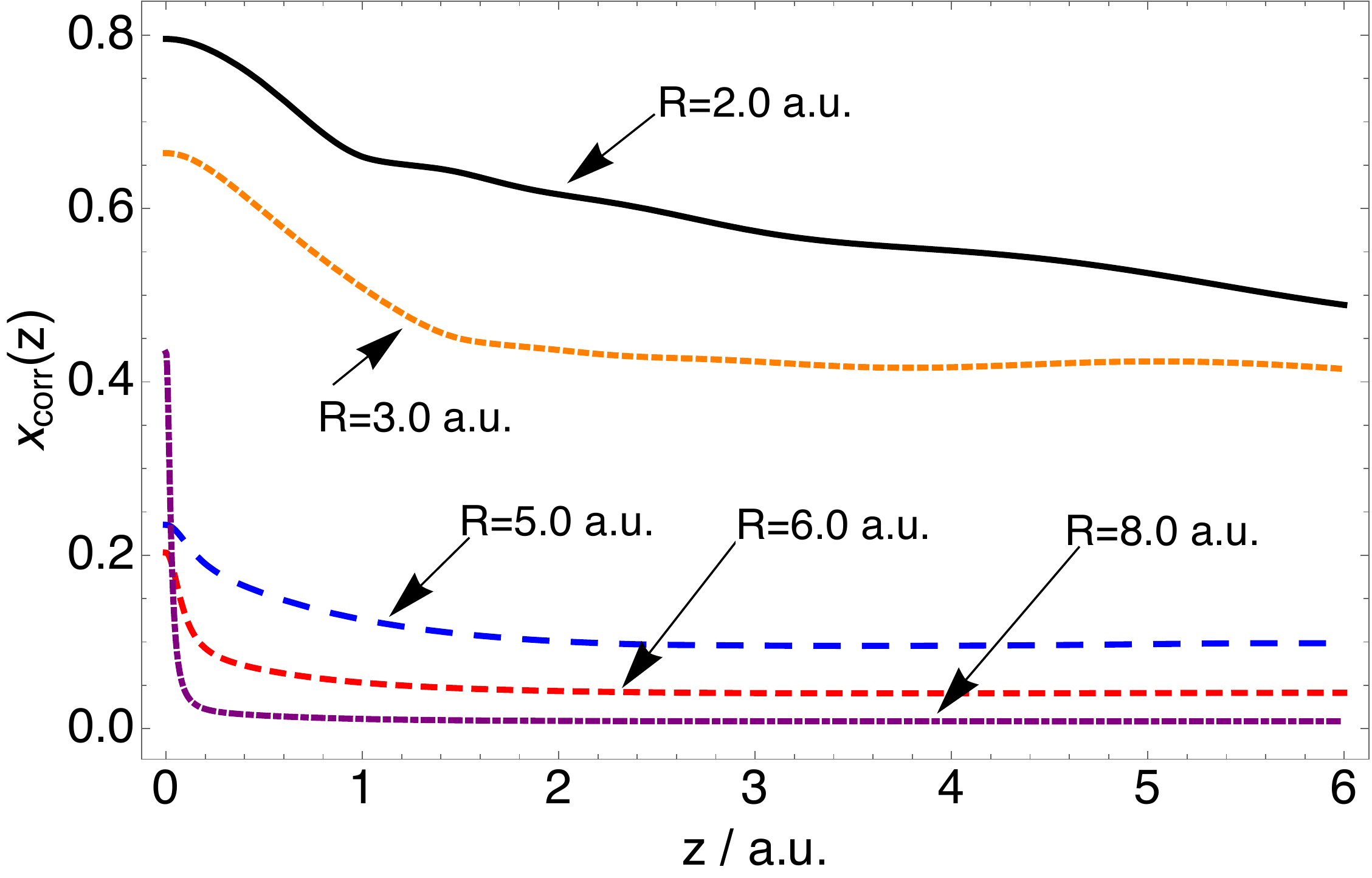}
\caption{$x_{\rm corr}(z)$  local correlation-type indicator of  eq~\ref{eq:xlam}.  shown for the H$_2$ molecule along the internuclear axis at different bond lengths, where z is the distance from the bond midpoint}
\label{fig_xlamloch2}
\end{figure}
%%%%%%%%%

In Figure~\ref{fig_xlamloch2} we show the $x_{\rm corr}(z)$ curves for the H$_2$ molecule along the internuclear axis for several bond lengths as a function of the distance from the bond midpoint, $z$. We can see that at the smaller bond length ($R=2.0$) the structure of the $x_{\rm corr}(z)$ curve is similar to the one of the helium atom. We also see that as we stretch the bond, the curves have a more linear structure. For the stretched H$_2$ we can see that {\em hot static correlation regions} are present at almost all points in space. Very small $x_{\rm corr}$ values indicate that in all these points we have the similar ``L-shaped'' local AC curves.\cite{TeaCorHel-JCP-09,locpaper} The exception is the energetically unimportant bond midpoint of stretched H$_2$, at which in case of infinitely stretched H$_2$ the local AC becomes: $w_{0 \leq \lambda \leq 1}(z)=w_0(z)$, with $x_{\rm corr}(z)=1$ and this happens because of the antisymmetry of the correlation hole at bond midpoint of H$_2$ at $R \to \infty$.\cite{BaeGri-JPCA-97}

%%%%%%%%%%
\begin{figure}
\includegraphics[width=\linewidth]{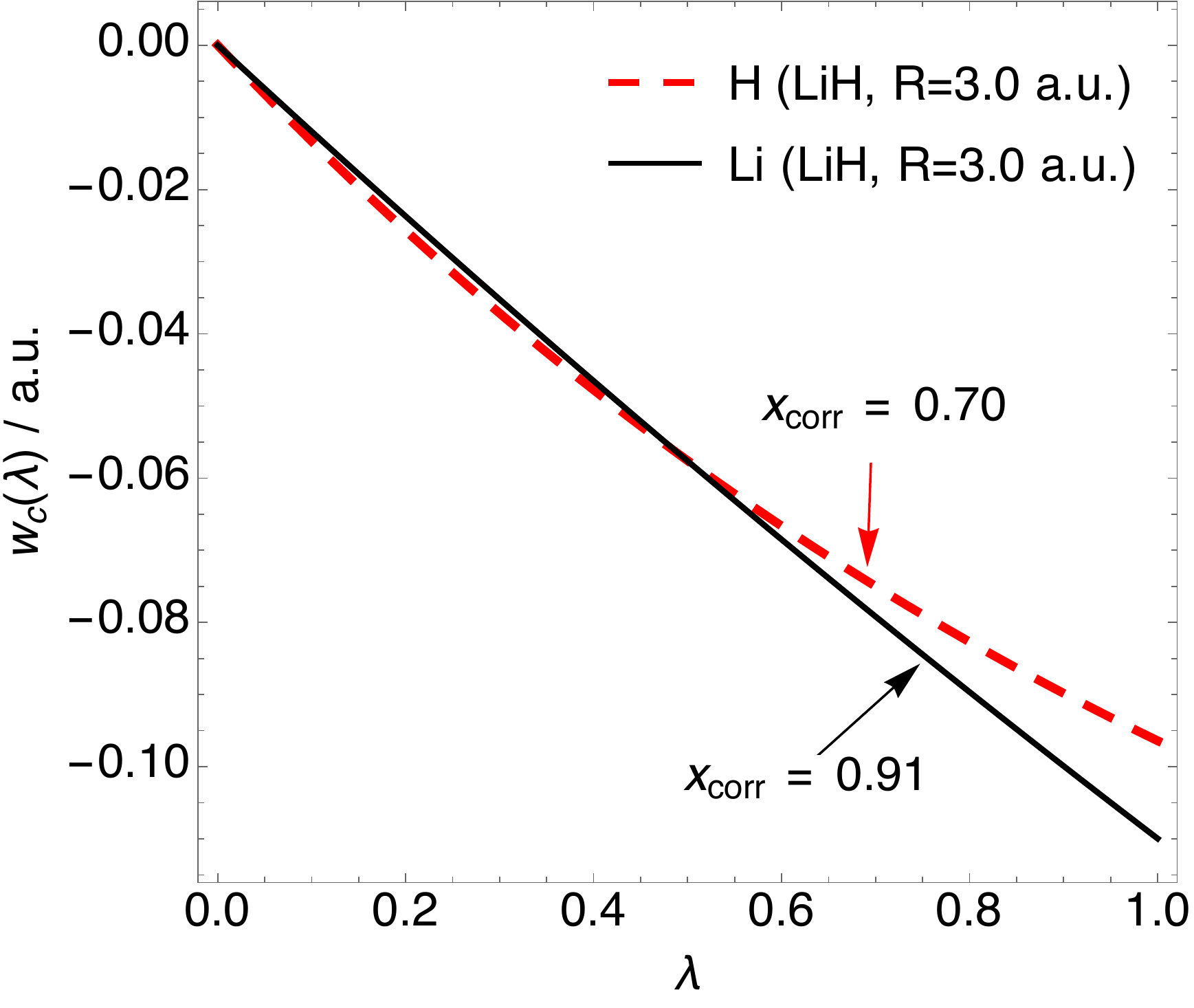}
\includegraphics[width=\linewidth]{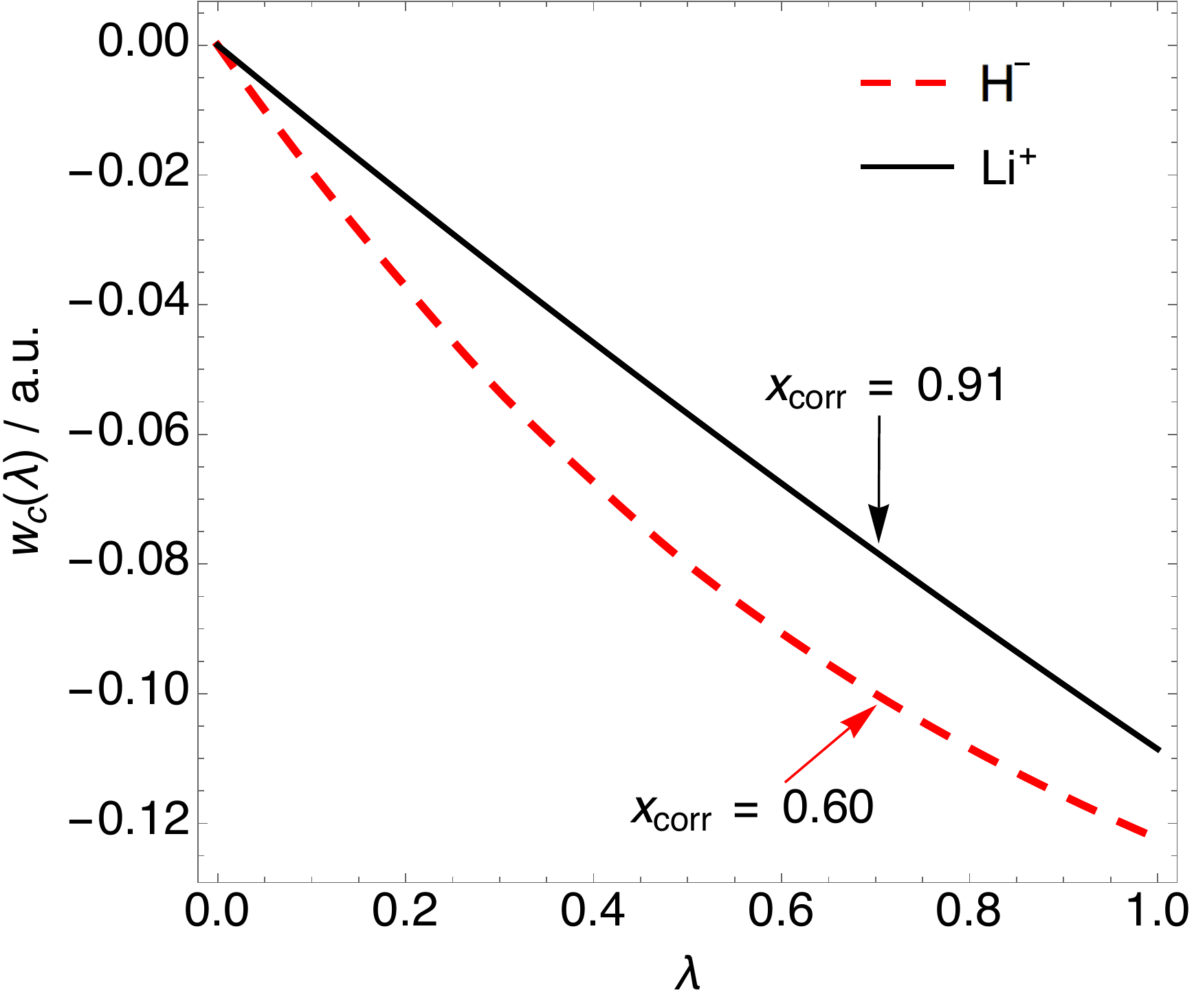}
\caption{The local correlation AC curves at the two nuclei of LiH at $R=3.0$~a.u (left panel) and at the nuclei of H$^-$ and Li$^{+}$ ions (right panel)} 
\label{fig_lihAC}
\end{figure}
%%%%%%%%%

In the case of LiH we observed an interesting difference between shapes of the local AC curves at the hydrogen and lithium nucleus. In the left panel of Figure~\ref{fig_lihAC} we show the correlation part of the local AC curves at both the hydrogen and lithium nuclei in near-equilibrium region, $R=3.0~a.u$. We can see that the curvature of the local AC curve at the hydrogen nucleus is much more pronounced than that of the lithium nucleus. As a result of the electronegativity difference between the lithium and hydrogen atom, we expect that LiH at equilibrium has a significant ionic character. As a result of the the bond polarization of LiH, we would expect that the hydrogen atom would have slightly anionic character and that the lithium atom would have slightly cationic character. That fact is mirrored by the observation of the corresponding local AC integrands and associated $x_{\rm corr}(\rv)$ values. From Figure~\ref{fig_lihAC} we can see that the local AC at the Li nucleus of LiH at $R=3.0~a.u.$ is nearly a straight line, as it is the case with the local AC at the nucleus of Li$^+$ ion (shown in the right panel of the same figure). The two corresponding $x_{\rm corr}(\rv)$ are the same and very close to $1$. On the other hand, the highly pronounced curvature present in the local AC at the H nucleus of LiH is very similar to that of the hydride ion. The $x_{\rm corr}(\rv)$ value at the H nucleus of LiH is $0.7$, indicating a significant presence of static correlation in this region, but still somewhat lower than at the nucleus of the hydride ion with $x_{\rm corr}(\rv)=0.6$.

Both $X_{\rm corr}$ and $x_{\rm corr}(\rv)$ quantities can be used in the context of the interpolation along the adiabatic connection. It might be the case that the certain interpolation forms are better suited for a particular correlation regime than for the others. For instance, we noticed that the SPL model works better than the others for atoms (with the usually high $X_{\rm corr}$ value), while the LB and two-legged interpolation performed better than the SPL for the intermediate correlation regimes (such as the H$_2$  at about $R=5.0~a.u$). We also see in this work that the interpolation can be even more sensitive to the interpolation form than the different input for the $\lambda \to \infty$ quantities. To tune the interpolation accuracy, a new XC functional can be constructed in which the correlation energies obtained by different interpolations (or the correlation energy densities in the local interpolation variant) are mixed linearly. The linear mixing parameters can depend on $X_{\rm corr}$ and thus be system--dependent: e.g. for large $X_{\rm corr}$ the total correlation would have a higher portion of the SPL interpolated correlation and  a larger portion of the correlation obtained by the LB and the two-legged representation interpolation for smaller $X_{\rm corr}$. We will try to pursue this idea in future work. 

\subsection{Lower bound to the exact energy}\label{low_bou}
In wave function theory (WFT), the energies obtained by variational methods are always upper bounds to the exact ground state energy. The larger the space for the trial wave function, the closer the energy to the true one is. This appealing feature is lost in KS DFT based on DFA, as the energies can be both higher and lower than the exact ones. It was shown in Ref~\citenum{MalGor-PRL-12} that if we approximate the AC integrand with the single line segment $\mathcal{W}_\lambda[\rho] = \mathcal{W}_\infty[\rho]$, we always obtain a lower bound to the exact energy. In this approximation, called KS SCE, the exchange-correlation functional is simply given by $E_{\rm xc}[\rho] = \mathcal{W}_\infty[\rho]$. The fact that the global AC curve is monotonically decreasing ensures that the KS SCE energies (both the self-consistent ones and the ones evaluated on the exact densities) are always lower than the exact ones. For systems in which static correlation dominates strongly, the KS SCE method gives reasonable energies\cite{MenMalGor-PRB-14,MalMirGieWagGor-PCCP-14,VucWagMirGor-JCTC-15}, tending towards the exact ones in the low-density limit. On the other hand, for systems where correlation is weak or moderate the KS SCE energies are too low.\cite{MalMirGieWagGor-PCCP-14,CheFriMen-JCTC-14,VucWagMirGor-JCTC-15} In these scenarios, the bound is very (sometimes extremely) loose. The KS SCE energies can be improved if we add corrections to them\cite{VucWagMirGor-JCTC-15} or if we use them as input in an interpolation scheme,\cite{locpaper}, a procedure we have also followed in the previous sections of this work. In this case, we can obtain energies that are substantially improved, but, as in other DFA's, they can be both higher and lower than the exact ones.

In this section we propose a way to tighten the lower bound given by the KS SCE energy for a given density. We do this by redefining the two-legged representation interpolation, using $\mathcal{W}_0[\rho]$, $\mathcal{W}'_0[\rho]$ and $\mathcal{W}_\infty[\rho]$ as input,
\begin{equation}\label{eq:2legsce}
    \mathcal{W}_\lambda = \begin{cases} 
        \mathcal{W}_0 + \lambda \mathcal{W}'_0 & \lambda \leqslant X_{\rm corr}^{\rm SCE} \\ 
        \mathcal{W}_\infty & \lambda > X_{\rm corr}^{\rm SCE}
    \end{cases} 
\end{equation}
where,
\begin{equation}\label{eq:xlamsce}
    X_{\rm corr}^{\rm SCE} = \frac{\mathcal{W}_\infty-\mathcal{W}_0}{\mathcal{W}'_0}.
\end{equation}

%\begin{align}
%W_\lambda= \left\{
%  \begin{array}{lr}
%   W_0+W'_0 \lambda , \quad \lambda \leqslant X_{\rm corr}^{\rm SCE}\\
%    W_\infty,\quad \lambda > X_{\rm corr}^{\rm SCE}
%  \end{array}
%\right.
%		\label{eq:2legsce}
%	\end{align}
%	where,	
%	\begin{align}
%X_{\rm corr}^{\rm SCE}=\frac{W_\infty-W_0}{W'_0}
%		\label{eq:xlamsce}
%	\end{align}
We call this interpolation 2-leg SCE interpolation. For $\mathcal{W}_\lambda$ of eq~\ref{eq:2legsce} to be a rigorous lower bound to the exact $\mathcal{W}_\lambda$, two conditions have to be satisfied. The first one is the monotonically decreasing nature of $\mathcal{W}_\lambda$, which is known to be true.\cite{LevPer-PRA-85,TeaCorHel-JCP-10} This condition ensures that ``the second leg'' of eq~\ref{eq:2legsce} is below the exact integrand curve: $\mathcal{W}_\infty \leq \mathcal{W}_\lambda $. The second condition is the convexity of the $\mathcal{W}_\lambda$ integrand. If $\mathcal{W}_\lambda$ as function of $\lambda$ is convex, then $\mathcal{W}_0+\mathcal{W}'_0 \lambda \leqslant \mathcal{W}_\lambda$. The convexity of $\mathcal{W}_\lambda$ is often assumed to be true, but it can actually be violated in the case of phase transitions along the adiabatic connection path. In these cases, the adiabatic connection curve would have jumps and would be only piecewise convex. For example, if our physical, $\lambda=1$, system is the uniform electron gas at a density lower than the one at which the ferromagnetic transition occurs, then the curve could have a small jump. In most cases, however, we might still expect, as it is often done, that the density constraint on a chemical system might prevent this from happening.
Besides being monotonically decreasing, it is also known that $\mathcal{W}_\lambda$ is bounded from below by the Lieb--Oxford inequality:\cite{LieOxf-IJQC-81}
\begin{equation}\label{eq:lo}
    \mathcal{W}_\lambda[\rho] \geq -C_{\rm LO} \int \rho^{4/3}\,{\rm d}\mathbf{r},
\end{equation}
%	\begin{align}
%W_\lambda[\rho] \geq -C_{\rm LO} \int \rho^{4/3} d\rv,
%		\label{eq:lo}
%	\end{align}
where $C_{\rm LO}$ is a constant rigorously known to be between $1.4119$\cite{SeiVucGor-MP-16} and $1.6358$.\cite{ChaHan-PRA-99} In addition to the 2-leg SCE lower bound, another lower bound can be constructed by replacing $\mathcal{W}_\infty[\rho]$ appearing in eq~\ref{eq:2legsce} with  $-C_{\rm LO} \int \rho^{4/3}\,{\rm d}\mathbf{r}$:
\begin{equation}\label{eq:2leglo}
    \mathcal{W}_\lambda = \begin{cases} 
        \mathcal{W}_0 + \lambda \mathcal{W}'_0 & \lambda \leqslant X_{\rm corr}^{\rm LO} \\ 
        -C_{\rm LO} \int \rho^{4/3}\,{\rm d}\mathbf{r} & \lambda > X_{\rm corr}^{\rm LO}
    \end{cases} 
\end{equation}
where,
\begin{equation}\label{eq:xlamlo}
    X_{\rm corr}^{\rm LO} = \frac{-C_{\rm LO} \int \rho^{4/3}\,{\rm d}\mathbf{r}-\mathcal{W}_0}{\mathcal{W}'_0}.
\end{equation}

%\begin{align}
%W_\lambda= \left\{
%  \begin{array}{lr}
%   W_0+W'_0 \lambda , \quad \lambda \leqslant X_{\rm corr}^{\rm LO}\\
%    -C_{\rm LO} \int \rho^{4/3} \mathrm{d}\rv,\quad \lambda > X_{\rm corr}^{\rm LO}
%  \end{array}
%\right.
%		\label{eq:2legsce}
%	\end{align}
%	where,	
%	\begin{align}
%X_{\rm corr}^{\rm LO}=\frac{-C_{\rm LO} \int \rho^{4/3} d\rv-W_0}{W'_0}
%		\label{eq:xlamsce}
%	\end{align}
We call this interpolation from the 2-leg LO interpolation. Since $\mathcal{W}_\infty[\rho] \leq -C_{\rm LO} \int \rho^{4/3} \,{\rm d}\mathbf{r}$ correlation energies from the 2-leg LO interpolation would be also lower bound to the exact correlation energies assuming convexity of the global AC integrand. In this work we use the following value for the LO constant: $C_{\rm LO}=1.4174$. The given LO constant was obtained by computing explicitly the indirect energies of the uniform electron spheres by using the SCE methodology, and extrapolating the value of $C_{\rm LO}$ in the $N \to \infty$ limit.\cite{SeiVucGor-MP-16} The value is very close to the lowest ever rigorously observed value for the given constant: $1.4119$.\cite{SeiVucGor-MP-16} This value is also lower than the value that was previously believed to be a lower bound for $C_{\rm LO}$, namely $1.444$, which obtained from the total energy of the bcc crystal of the uniform electron gas. Lewin and Lieb have recently shown that this value does not correspond to an indirect energy.\cite{LewLie-PRA-15}

%%%%%%%%%%%%%%%%%%%%%%%%%%%%%%%%%%%%%%%%%%%%%%%%%%%%%%%%%%%%%%%%%%%%%%%%%%%%
%%%%%%%%%%%%%%%%%%%%%%%%%%%%%%%%%%%%%%%%%%%%%%%%%%
%\begin{table}[]
%\centering
%\caption{}
%\label{tab_legsce}
%\begin{tabular}{llllll}
%system       & reference & KS SCE  & LO      & 2-leg SCE & 2-leg LO \\
%H$^-$         & -0.0409   & -0.1860  & -0.2662 & -0.0571   & -0.0571  \\
%He            & -0.0400     & -0.4733 & -0.6689 & -0.045    & -0.0450   \\
%Be            & -0.0920    & -1.3464 & -1.7775 & -0.1234   & -0.1234  \\
%Ne$^6+$       & -0.1833   & -3.9666 & -5.1464 & -0.2779   & -0.2779  \\
%Ne            & -0.347    & -7.916  & -9.047  & -0.437    & -0.437   \\
%Ar            & -0.404    & -20.944 & -23.272 & -0.511    & -0.511   \\
%H$_2$ $R=1.4$ & -0.04     & -0.302  & -0.431  & -0.049    & -0.049   \\
%H$_2$ $R=2.8$ & -0.068    & -0.225  & -0.358  & -0.102    & -0.102   \\
%H$_2$ $R=5.0$ & -0.184    & -0.234  & -0.401  & -0.221    & -0.361   \\
%H$_2$ $R=7.0$ & -0.234    & -0.25   & -0.432  & -0.248    & -0.425   \\
%H$_2$ $R=9.0$ & -0.256    & -0.262  & -0.448  & -0.262    & -0.446  
%\end{tabular}
%\end{table}

\begin{table}
\caption{The correlation energy obtained from the KS--SCE model, LO bound, 2-leg SCE model and 2-leg LO model compared with the reference value.} \label{tab_legsce}
\centering
\begin{tabular*}{\linewidth}{l@{\extracolsep{\fill}}lllll} \hline\hline\noalign{\vskip 1ex}
Species       & $E_c^{\rm ref}$ & KS SCE  & LO      & 2-leg SCE & 2-leg LO \\[0.5ex]\hline\noalign{\vskip 0.5ex}
H$^-$         & -0.0409         & -0.1860 & -0.2662 & -0.0571   & -0.0571  \\
He            & -0.0400         & -0.4733 & -0.6689 & -0.0450   & -0.0450  \\
Be            & -0.0920         & -1.3464 & -1.7775 & -0.1234   & -0.1234  \\
Ne$^6+$       & -0.1833         & -3.9666 & -5.1464 & -0.2779   & -0.2779  \\
Ne            & -0.3470         & -7.9160 & -9.0470 & -0.4370   & -0.4370  \\
Ar            & -0.4040         & -20.944 & -23.272 & -0.5110   & -0.5110  \\
H$_2$ $R=1.4$ & -0.0400         & -0.3020 & -0.4310 & -0.0490   & -0.0490  \\
H$_2$ $R=2.8$ & -0.0680         & -0.2250 & -0.3580 & -0.1020   & -0.1020  \\
H$_2$ $R=5.0$ & -0.1840         & -0.2340 & -0.4010 & -0.2210   & -0.3610  \\
H$_2$ $R=7.0$ & -0.2340         & -0.2500 & -0.4320 & -0.2480   & -0.4250  \\
H$_2$ $R=9.0$ & -0.2560         & -0.2620 & -0.4480 & -0.2620   & -0.4460  \\[1ex]\hline\hline
\end{tabular*}
\end{table}
%%%%%%%%%%%%%%%%%%%%%%%%%%%%%%%%%%%%%%%%%%%%%%%%%%%%%%%%%%%%%%%%%%%%%%%%%%%%
%%%%%%%%%%%%%%%%%%%%%%%%%%%%%%%%%%%%%%%%%%%%%%%%%%

In Table~\ref{tab_legsce} we show the reference correlation energies (CCSD), the KS SCE ones (i.e,  $\mathcal{W}_\infty[\rho]-\mathcal{W}_0[\rho]$), the correlation energies obtained from the LO bound  (i.e., $-C_{\rm LO} \int \rho^{4/3}\,{\rm d}\mathbf{r} - \mathcal{W}_0[\rho]$), together with the correlation energies obtained by the 2-leg SCE and 2-leg LO interpolations. First of all, we immediately see that already the KS SCE correlation energies are significantly above the ``LO correlation energies''. We also see from this table that the 2-leg SCE substantially tightens the lower bound  of KS SCE, even by an order of magnitude in some cases. Notice that the accuracy of the 2-leg SCE is not as high as the accuracy of the other interpolations presented in this work, but the main advantage of it is that it recovers the KS SCE feature to give correlation energies that are a lower bound to the exact ones. When $X_{\rm corr}^{\rm SCE} \geqslant 1$, the correlation energy obtained from the interpolated $\mathcal{W}_\lambda[\rho]$ of eq~\ref{eq:2legsce} becomes $\mathcal{W}_0'[\rho]/2=E_c^{\rm GL2}[\rho]$. This happens for the atoms given in Table~\ref{tab_legsce} and this is why both 2-leg SCE and 2-leg LO give the same correlation energies for the given atoms. However, the role of $\mathcal{W}_\infty[\rho]$ in the 2-leg SCE interpolation is to correct $E_c^{\rm GL2}[\rho]$ energy when it becomes too low. This happens for example in the stretched H$_2$, e.g. at $R=10.0~a.u.$ (where $X_{\rm corr}^{\rm SCE} << 1$) the GL2 correlation energy is too low, $E_c^{\rm GL2}[\rho] \sim -80~E_h$, but the 2-leg SCE interpolation gives a value very close to the exact one. We can see how the 2-leg SCE and the other interpolations employed here can benefit from the complementary information provided by the $\mathcal{W}_\infty[\rho]$ and $\mathcal{W}_0'[\rho]$ quantities. For the stretched H$_2$ we can see that the 2-leg SCE correlation energies are much closer to the exact ones than the ones of the 2-leg LO interpolation. The latter could be tightened if instead of using $1.41$, we use $C=1.21$, which should be the optimal constant for $N=2$,\cite{SeiVucGor-MP-16} but that would lead to the loss of generality w.r.t number of electrons. 
 
%%%%%%%%%%%%%%%%%%%%%%%%%%%%%%%%%%%%%%%%%%%%%%%%%%%%%%%%%%%%%%%%%%%%%%%%%%%%%%%%%
\begin{figure}
\includegraphics[width=\linewidth]{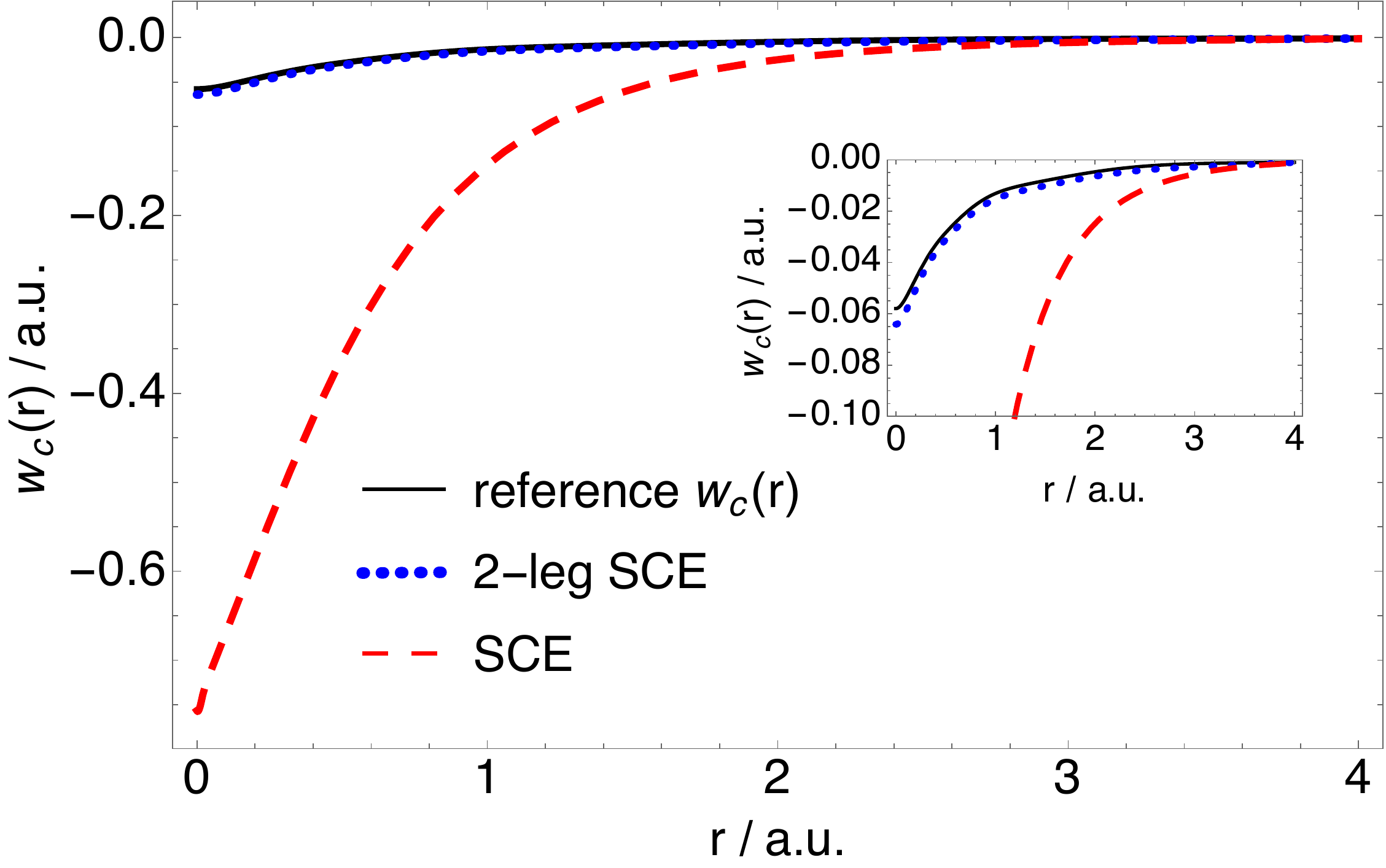}\\
\includegraphics[width=\linewidth]{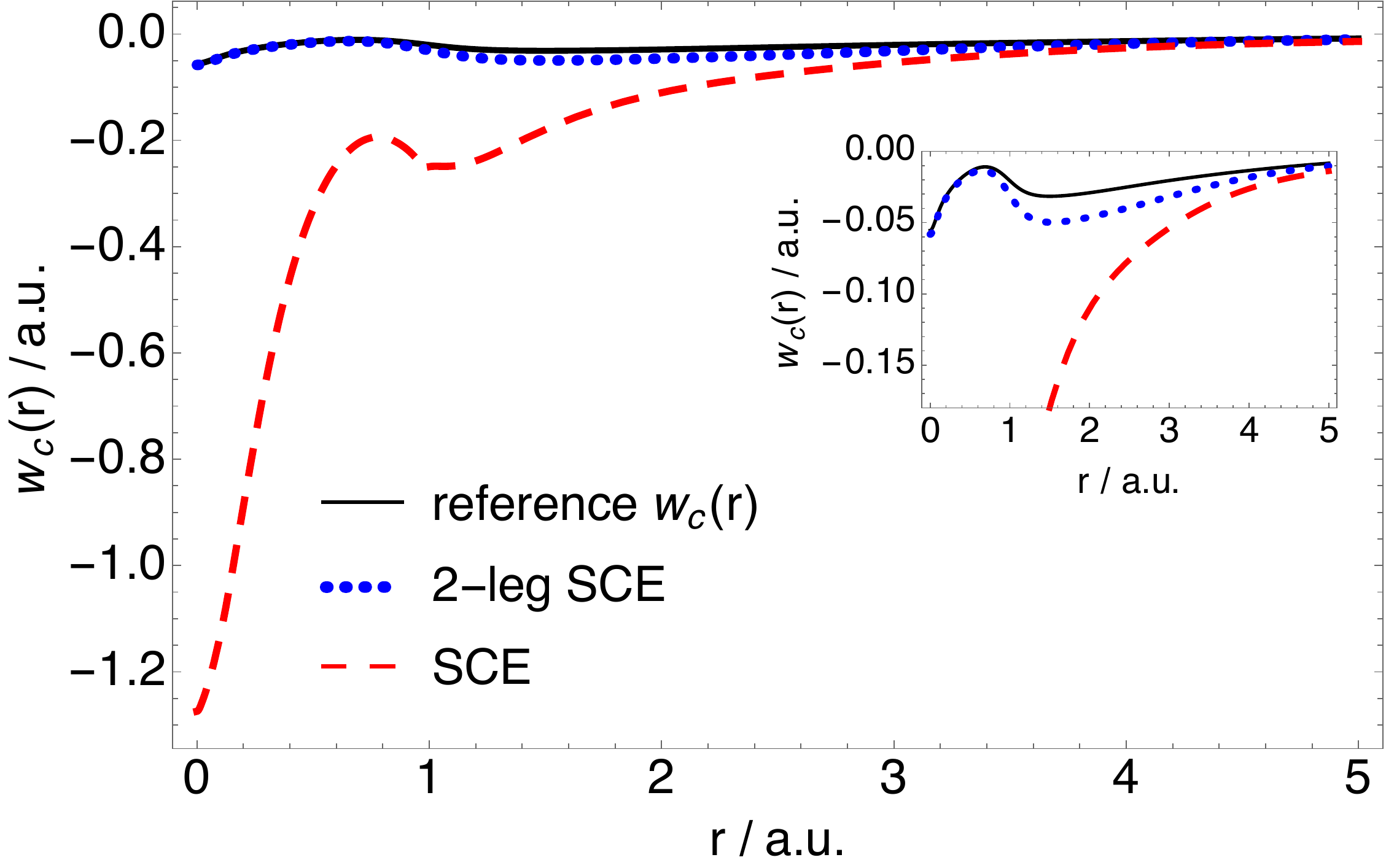}
\caption{The coupling constant averaged correlation energy density ($\bar w_c(\rv)$) obtained by the 2-leg SCE interpolation (the local variant of eq~\ref{eq:2legsce}) shown together with the reference $\bar w_c(\rv)$ and the SCE correlation energy density ($\bar w_\infty^{c}(\rv)$. Upper panel: He atom (reference $\bar w_c(\rv)$ obtained by FCI), lower panel: Be atom (reference $\bar w_c(\rv)$ obtained by CCSD)} 
\label{fig_w_leg_atoms}
\end{figure}
%%%%%%%%%%%%%%%%%%%%%%%%%%%%%%%%%%%%%%%%%%%%%%%%%%%%%%%%%%%%%%%%%%%%%%%%%%%%%

It would be interesting to see whether the energy densities obtained by the local variant of eq~\ref{eq:2legsce} give a lower bound to $w_\lambda(\rv)$ and thus also a lower bound to $\bar{w}_{xc}(\mathbf{r})$. For this to be true, we two conditions should again be satisfied: the monotonic decrease and convexity of $w_\lambda(\rv)$ with respect to $\lambda$. Our numerical results for Coulombic systems suggest that both conditions are satisfied.\cite{IroTea-MP-15,locpaper} The counterexample to the monotonicity of $w_\lambda(\rv)$ is known for the systems in which the external potential is not Coulombic, i.e, in the Hooke's atom series, where in the energetically unimportant region (in the density tail), $w_\infty(\rv)$ can be above $w_1(\rv)$.\cite{MirSeiGor-JCTC-12} In Figure~\ref{fig_w_leg_atoms} we show the coupling constant averaged correlation energy $\bar w_c(\rv)$  obtained by the local variant of the 2-leg SCE interpolation, together with the reference $\bar w_c(\rv)$ and the $w_c(\rv)$ corresponding to the $\lambda \to \infty$ (SCE) limit, $w_\infty(\rv)-w_0(\rv)$, for the He atom (left panel) and the Be atom (right panel). In both cases, the SCE correlation energy densities are a lower bound to the reference $\bar w_c(\rv)$, but we can see that this bound is very loose, especially in the region near the nuclei. With the 2-leg SCE local interpolation method we can see that the local bound has been substantially tightened. We do not attempt here to construct a local variant of the 2-leg LO bound, as the rigorous local variant of the LO inequality that binds the energy density in the gauge of the XC hole is not known.\cite{MirSeiGor-JCTC-12}

\section{Conclusions and Perspectives}
\label{sec_concl}

Interpolating {\em locally} along the adiabatic connection between the weak and strong coupling limits has manifold advantages: unlike previous efforts in this direction based on global (integrated over all space) adiabatic connection models, it does not violate size-consistency, at least in the absence of degeneracy. While most of the present density functional approximations have a bias towards weak correlation, the inclusion of information from the strictly correlated electrons limit leads to a more balanced approach, avoiding bias towards a particular correlation regime. This approach does not suffer from the exchange energy density gauge problem as a result of the compatibility of the energy densities at strong correlation with the exact exchange energy density. The main focus of this work was to test how the replacement of the comutationally expensive SCE energy densities with approximate strong-coupling energy densities affects the local interpolation scheme. For this purpose, we employed different models for the SCE energy densities in the gauge of the XC hole: the nonlocal radius functional (NLR) (a functional which retains some of the SCE nonlocality) and the (semi)local ``point-charge plus continuum''(PC) model functional. We used atomic correlation energies, together with the hydrogen molecule and the lithium hydride dissociation curves, as simple tests. For these systems all the other ingredients (exchange energy densities and local slopes) are available to high accuracy, which allows us to isolate the effect of the error coming from the approximations for the strong-coupling limit.

These tests showed that the NLR energy densities are an excellent alternative to the SCE energy densities for the local interpolation. The energy densities with the PC model are very easy to obtain, but the overall performance of the local interpolation based on the PC model was not satisfactory. These interpolations are adequate for the atomic correlation energies but they introduce a very large error for the considered molecular dissociation curves. As the error in the PC model based local interpolation is already too large, the global interpolation based on the PC-GGA model seems to be more promising despite the size consistency issues with the global interpolations. Extensive testing of the PC model in the context of the global interpolation has been very recently carried on in Ref.~\onlinecite{FabGorSeiSal-JCTC-16}, where some of the limitations of global interpolations have been carefully studied. 

We have also used interpolation models along the adiabatic connection to propose a correlation-type indicator and a tighter lower bound to the exact XC energy. 
In our future work we will try to use these two quantities to improve the accuracy of the local interpolations. We will also test the recently proposed model of Bahmann, Zhou, and Ernzerhof,\cite{BahZhoErn-JCP-16} which should provide, in principle, even better results than the original NLR approximation.

\section*{Acknowledgements}\label{ack}
We acknowledge financial support from the European Research Council under H2020/ERC Consolidator Grant corr-DFT (Grant No. 648932) and the Netherlands Organization for Scientific Research (NWO) through an ECHO grant (717.013.004). A. M. T. is grateful for support from the Royal Society University Research Fellowship scheme. A. M. T and T. J. P. I. are grateful for support from the Engineering and Physical Sciences Research Council (EPSRC), (Grant No. EP/M029131/1). We are grateful for access to the University of Nottingham High Performance Computing Facility.

%%%%%%%%%%%%%%%%%%%%%%%%%%%%%%%%%%%%%%%%%%%%%%%%%%%%%%%%%%%%%%%%%%%%%%%%%%%%%%%%
 \appendix
 \section{The mathematical forms of the interpolation models for the adiabatic connection}\label{app_interpol}

%%%%%%%%%%%%%%%%%%%%%%%%%%%%%%%%%%%%%%%%%%%%%%%%%%%%%%%%%%%%%%%%%%%%%%%%%%%%%%%%%
\begin{table*}
\caption{Forms of the adiabatic connection interpolation models (for the Pad{\'e}$[1/1]$ model, $p > 0$, $p \in \mathbb{R}$).}\label{tab_ac_eqns}
\centering
\begin{tabular*}{0.9\textwidth}{l@{\extracolsep{\fill}}mmmmc} \hline\hline\noalign{\vskip 1ex}
  & w_{\lambda}(\mathbf{r}) & a(\mathbf{r}) & b(\mathbf{r}) & c(\mathbf{r}) & Refs. \\[1ex]\hline\noalign{\vskip 1ex}
\textbf{SPL} & a + \frac{b}{\sqrt{1 + c \lambda}} & w_{\infty}(\mathbf{r}) & w_{0}(\mathbf{r}) - w_{\infty}(\mathbf{r}) & -\frac{2 w_{0}'(\mathbf{r})}{w_{0}(\mathbf{r}) - w_{\infty}(\mathbf{r})} & \citenum{locpaper,SeiPerLev-PRA-99,SeiGorSav-PRA-07} \\[4ex]
\textbf{LB} & a + b \left(\frac{1}{(1 + c \lambda)^{2}} + \frac{1}{\sqrt{1 + c \lambda}}\right) & w_{\infty}(\mathbf{r}) & (w_{0}(\mathbf{r}) - w_{\infty}(\mathbf{r}))/2 & -\frac{4 w_{0}'(\mathbf{r})}{5(w_{0}(\mathbf{r}) - w_{\infty}(\mathbf{r}))} & \citenum{locpaper,LiuBur-PRA-09} \\[4ex]
\textbf{Pad\'e}[1/1] & a + \frac{b \lambda}{1 + c \lambda} & w_{0}(\mathbf{r}) &  w_{0}'(\mathbf{r}) & \frac{-w_{0}(\mathbf{r}) + w_{p}(\mathbf{r}) - w_{0}'(\mathbf{r})}{w_{0}(\mathbf{r}) - w_{p}(\mathbf{r})} & \citenum{locpaper,Ern-CPL-96,SanCohYan-JCP-06} \\[2ex]\hline\hline
\end{tabular*}
\end{table*}
%%%%%%%%%%%%%%%%%%%%%%%%%%%%%%%%%%%%%%%%%%%%%%%%%%%%%%%%%%%%%%%%%%%%%%%%%%%%%%%%%

\bibliography{biblioPaola,biblio_spec,biblio1,biblio_add,biblio_add2}

\end{document}